\documentclass[useAMS, usegraphicx, usenatbib]{mn2e}
\usepackage{aas_macros}
\usepackage{amsmath}
\usepackage{graphicx}
\usepackage{natbib}
\usepackage{multirow}
\newcommand{\kunit}{\,h\,\mathrm{Mpc}^{-1}}
\newcommand{\lunit}{\,h^{-1}\,\mathrm{Mpc}}
\newcommand{\lunitk}{\,h^{-1}\,\mathrm{kpc}}
\newcommand{\munit}{\,h^{-1}\,\mathrm{M}_{\sun}}
\newcommand{\vunit}{\,\mathrm{km}\,\mathrm{s}^{-1}}
\setcitestyle{authoryear,round,semicolon,aysep={},yysep={,},notesep={}}
\title[Galaxy formation and the matter power spectrum]{The effects of galaxy formation on the matter power spectrum: A~challenge for precision
cosmology}
\author[M. P. van Daalen et al.]{Marcel P. van Daalen$^{1,2}$\thanks{E-mail: daalen@strw.leidenuniv.nl}, Joop~Schaye$^{1}$, C.~M.~Booth$^{1}$ \& Claudio~Dalla~Vecchia$^{1,3}$\\
$^1$Leiden Observatory, Leiden University, P.O. Box 9513, 2300 RA Leiden, The Netherlands\\
$^2$Max Planck Institute for Astrophysics, Karl-Schwarzschild Stra\ss{}e 1, 85741 Garching, Germany\\
$^3$Max Planck Institute for Extraterrestrial Physics, Giessenbachstra\ss{}e 1, 85748 Garching, Germany}
\begin{document}
\date{Accepted 2011 April 29. Received 2011 April 29; in original form 2011 March 16}
\pagerange{\pageref{firstpage}--\pageref{lastpage}} \pubyear{2011}
\maketitle
\label{firstpage}
\begin{abstract}
Upcoming weak lensing surveys, such as LSST, EUCLID, and WFIRST, aim to measure the matter power spectrum with unprecedented accuracy. In order to fully exploit these observations, models are needed that, given a set of cosmological parameters, can predict the non-linear matter power spectrum at the level of $1\%$ or better for scales corresponding to comoving wave numbers $0.1 \la k \la 10\kunit$. We have employed the large suite of simulations from the OWLS project to investigate the effects of various baryonic processes on the matter power spectrum. In addition, we have examined the distribution of power over different mass components, the back-reaction of the baryons on the CDM, and the evolution of the dominant effects on the matter power spectrum. 
We find that single baryonic processes are capable of changing the power spectrum by up to several tens of per cent. Our simulation that includes AGN feedback, which we consider to be our most realistic simulation as, unlike those used in previous studies, it has been shown to solve the overcooling problem and to reproduce optical and X-ray observations of groups of galaxies, predicts a decrease in power relative to a dark matter only simulation ranging, at $z=0$, from $1\%$ at $k \approx 0.3\kunit$ to $10\%$ at $k \approx 1\kunit$ and to $30\%$ at $k \approx 10\kunit$. This contradicts the naive view that baryons raise the power through cooling, which is the dominant effect only for $k \ga 70\kunit$. Therefore, baryons, and particularly AGN feedback, cannot be ignored in theoretical power spectra for $k \ga 0.3\kunit$. It will thus be necessary to improve our understanding of feedback processes in galaxy formation, or at least to constrain them through auxiliary observations, before we can fulfil the goals of upcoming weak lensing surveys. 
\end{abstract}

\section{Introduction}
One of the aims of cosmology is to find the initial conditions for structure formation in the Universe. These can be characterised by a single set of cosmological parameters, which directly influence the formation, growth and clustering of structure, and hence the distribution of matter as we observe it today. 

A powerful measure of the statistical distribution of matter (and a sufficient one for the case of Gaussian fluctuations), is the matter power spectrum, $P(k)$, where $k$ is the comoving wave number corresponding to a comoving spatial scale $\lambda=2\pi/k$. Given a sufficiently accurate model for the formation of structure, we can infer the initial, linear power spectrum from the observed, non-linear one. Moreover, as the rate of growth of structure depends on the expansion history, such a model also allows us to convert observations of the evolution of the power spectrum into measurements of other cosmological parameters such as the equation of state of the dark energy. 

Some of the most accurate measurements of the matter power spectrum come from studies of weak, gravitational lensing \citep[e.g.][]{Massey2007, Fu2008, Schrabback2010}, galaxy clustering \citep[e.g.][]{Cole2005, Reid2010} and the Ly$\alpha$ forest \citep[e.g.][]{Viel2004, McDonald2006}. Up to a few years ago, the statistical errors were sufficiently large that one could use analytical predictions (always assuming, amongst other things, that the Universe contains only dark matter), such as those by \citet{PeacockDodds1996}, \citet{Ma1999} and \citet{SmithPeacock2003}. The latter used ideas from the ``halo model'' \citep[e.g.][]{PeacockSmith2000, Seljak2000, CooraySheth2002} to improve upon the accuracy of simpler analytical predictions. In recent years the further improvement of this model has become increasingly dependent on the results from $N$-body simulations, such as the derived concentration-mass relation for dark matter haloes \citep[e.g.][]{Neto2007, Duffy2008, Hilbert2009}. If baryonic effects were negligible, then these methods would allow the matter power spectrum to be predicted with an accuracy of $\sim 1\%$ for wave numbers $k \la 1\kunit$ \citep{Heitmann2010}. However, we will show here that baryonic effects are larger than this on the scales relevant for many observations.

Upcoming weak lensing surveys aim to measure the matter power spectrum on scales of $0.1\kunit < k < 10\kunit$. In order to reach the level of precision their instruments are capable of, surveys such as LSST,\footnote{\texttt{http://www.lsst.org/lsst}} EUCLID,\footnote{\texttt{http://www.euclid-imaging.net/}} and WFIRST\footnote{\texttt{http://wfirst.gsfc.nasa.gov/}} need to be calibrated using theoretical models that retain $1\%$ accuracies on these scales \citep{HutererTakada2005, Laureijs2009}.\footnote{Since cosmological parameters are inferred from cosmic shear using a complicated weighting of the power spectrum over a range of scales and redshifts, the relation between the accuracy with which these parameters can be determined and the uncertainty in the models depends on the survey and is different for different parameters. Semboloni et al.\ (in preparation) will present a more detailed study of the consequences of our findings for weak lensing surveys.} This is, however, not as straightforward as increasing the resolution of existing $N$-body simulations: many authors have demonstrated that on these scales baryonic matter, which is not accounted for in currently employed theoretical models, introduces deviations of up to $10\%$ \citep{White2004, ZhanKnox2004, Jing2006, Rudd2008, Guillet2009, Casarini2010}. 

Recent hydrodynamic simulations include many of the physical processes associated with baryons, such as radiative cooling, star formation and supernova (SN) feedback. However, the processes which cannot be resolved in simulations are generally also not entirely understood, and different prescriptions exist that aim to model the same physics. Because of this, different authors may find significantly different results even when including the same baryonic processes. Furthermore, it is not a priori clear which physical effects are capable of changing the matter power spectrum at the $1\%$ level and should therefore be included. These modelling uncertainties may thus prevent upcoming surveys from further constraining the cosmological parameters of our Universe.

Here we employ a large suite of state-of-the-art cosmological, hydrodynamical simulations from the OWLS project \citep{Schaye2010} to systematically study the effects of various baryonic processes on the matter power spectrum over a wide range of scales, $k \sim 0.1-500\kunit$. These processes include metal-line cooling, different prescriptions for SN feedback, and feedback from active galactic nuclei (AGN). We will see that all of our results are heavily influenced by the inclusion of AGN feedback, which was not considered by earlier studies and which has been shown to solve the overcooling problem that has long plagued hydrodynamical simulations and to lead to an excellent match to both the optical and X-ray properties of groups of galaxies \citep{McCarthy2010,McCarthy2011}. Outflows driven by AGN strongly increase the scale out to which baryons modify the power spectrum. We also investigate how the power is distributed over different components (i.e.\ CDM, gas and stars) and examine the back-reaction of the baryons on the dark matter. In a follow-up paper (Semboloni et al., in preparation), we will quantify the implications for current and proposed weak lensing surveys and we will show how the uncertainty due to baryonic physics can be reduced by making use of additional observations of groups and clusters.

This paper is organised as follows. In \S\ref{sec:simulations} we discuss the simulations and the power spectrum estimator employed. In our main results section, \S\ref{sec:results}, we compare our dark matter only simulation to analytical estimates (\S\ref{sec:dmonly}), we compare power spectra of simulations with different baryonic processes (\S\ref{sec:barprocesses}), and we investigate how the power is distributed over different physical components (\S\ref{sec:contributions}). In this section we also examine the back-reaction of galaxy formation on the dark matter (\S\ref{sec:backreaction}) and we consider the evolution of the most dominant effects on the power spectrum (\S\ref{sec:agn}). We compare to the results found by other authors in \S\ref{sec:previouswork} and provide a summary in \S\ref{sec:conclusions}. Finally, we test the convergence of our results in Appendix~A and provide tables of the power spectra of all simulations in Appendix~B.

We note that all distances quoted in this paper are comoving and all power spectra are obtained at redshift zero, unless stated otherwise.

\section{Simulations}
\label{sec:simulations}
The OWLS project \citep{Schaye2010}, where OWLS is an acronym for OverWhelmingly Large Simulations, is a suite of large, cosmological, hydrodynamical simulations. The code used is a heavily extended version of \textsc{gadget iii}, a Lagrangian code which was last described in \citet{Springel2005}. It uses a TreePM algorithm to efficiently calculate the gravitational forces (where PM stands for Particle Mesh and the ``Tree'' describes the structure in which the particles are organised for this calculation, see for example \citealp{BarnesHut1986, Xu1995, Bagla2002}) and Smoothed Particle Hydrodynamics (SPH) to follow and evolve the gas particles \citep[see][ for a review]{Rosswog2009}.

There are two main sets of simulations, which have periodic boxes of size $L=25$ and $100\,h^{-1}$ comoving $\mathrm{Mpc}$ on a side, and are run down to redshifts $z=2$ and 0, respectively. Most simulations use $512^3$ collisionless cold dark matter (CDM) particles and an equal number of baryonic (collisional gas or collisionless star) particles. We will refer to the particle number used in a simulation with the parameter $N=N_\mathrm{part}^{1/3}$ ($=512$ for the high-resolution simulations). In this work we will focus on $z=0$ and hence on the simulations using a $100\lunit$ box. The particle masses are $4.06 \times 10^8\munit [L/(100\lunit)]^3 [N/512]^{-3}$ for the dark matter and $8.66 \times 10^7\munit [L/(100\lunit)]^3 [N/512]^{-3}$ for the baryons. The gravitational forces are softened on a comoving scale of $1/25$ of the initial mean interparticle spacing, $L/N$, but the softening length is limited to a maximum physical scale of $2\lunitk [L/(100\lunit)]$ which is reached at $z=2.91$. The SPH calculations use 48 neighbours. 

For the initial conditions, a theoretical matter power spectrum -- which of course depends on the chosen set of cosmological parameters -- is generated using \textsc{cmbfast} \citep[][, version 4.1]{SeljakZaldarriaga1996}. Prior to imposing the linear input spectrum, the particles are set up in an initially glass-like state, as described in \citet{White1994}. The particles are then evolved to redshift $z=127$ using the \citet{Zel'dovich1970} approximation.

On small scales, the physics of galaxy formation is unresolved, and subgrid models are needed to include baryonic effects like radiative cooling, star formation and supernova feedback. Although each OWLS run is a state-of-the-art cosmological simulation in itself, the real power of the OWLS project lies in the fact that it is composed of more than 50 simulations that all incorporate different sets of physical processes, parameter values, or subgrid recipes. In this way the effects of turning off or tweaking a single process can be studied in detail, making it especially well-suited to investigate which processes can, by themselves, change the power at $k\sim 1-10\kunit$ by $>1\%$. In this paper we briefly describe the subgrid physics included in the reference simulation, as well as the differences with respect to simulations we compare to in \S\ref{sec:barprocesses}. For a more detailed treatment of the simulations and the different physics models included, we refer to \citet{Schaye2010}.

\subsection{The reference simulation}
\label{sec:ref}
As the intention of the OWLS project is to investigate the effects of altering or adding a single physical process, it is convenient to have a single simulation that acts as the basis for all other simulations. Such a ``default'' simulation should of course include many of the physical processes that we know to be important already, as ideally we would only want to vary one thing at a time. We call this simulation the reference simulation, or \textit{REF} for short. Note that this is not intended to be the ``best'' simulation, but simply a model to build on. In fact, it has for example been shown that AGN feedback, which was not included in the \textit{REF} model and which we briefly discuss in the next section, is required to match observations of groups and clusters of galaxies \citep{McCarthy2010, McCarthy2011}.

We assume cosmological parameter values derived from the Wilkinson Microwave Anisotropy Probe (WMAP) 3-year	results	\citep{Spergel2007}: \{$\Omega_\mathrm{m}$, $\Omega_\mathrm{b}$, $\Omega_\mathrm{\Lambda}$, $\sigma_8$, $n_\mathrm{s}$, $h$\} = \{$0.238$, $0.0418$, $0.762$, $0.74$, $0.951$, $0.73$\}. Except for $\sigma_8$, all of these are consistent with the WMAP 7-year data \citep{Komatsu2011}. This specific parameter describes the root mean square fluctuation in spheres with a radius of $8\lunit$ linearly extrapolated to $z=0$ and effectively normalises the matter power spectrum. Measurements in the last few years have systematically increased the value of $\sigma_8$, which may influence the validity of our results. To check the effects of using ``wrong'' values for this and other cosmological parameters, we have re-run our two most important simulations -- one with only dark matter and one in which AGN feedback is added to the reference model -- using the WMAP7 cosmology. We briefly discuss these at the end of section \S\ref{sec:othersims}. As we shall see in \S\ref{sec:agn}, this change in cosmology does not affect our conclusions. 

The reference simulation includes both radiative cooling and heating, which are modelled using the prescription of \citet{Wiersma2009a}. Net radiative cooling rates are computed on an element-by-element basis in the presence of the cosmic microwave background and the \citet{HaardtMadau2001} model for the UV and X-ray background radiation from quasars and galaxies, taking into account the contributions of eleven different elements pre-computed using the publicly available photo-ionization package CLOUDY, last described by \citet{Ferland1998}. The effects of hydrogen ionization are modelled by switching on the \citet{HaardtMadau2001} model at $z=9$.

Cosmological simulations do not yet come close to resolving the process of star formation, and so a subgrid recipe has to be included for this as well. In our simulations, gas particles can be converted into star particles once their hydrogen number densities exceed the threshold for thermo-gravitational instability ($n_\mathrm{H}^*=0.1\,\mathrm{cm}^{-3}$; \citealp{Schaye2004}). Cold gas particles with higher densities follow an imposed equation of state, $P \propto \rho^{\gamma_\mathrm{eff}}$. Here $\gamma_\mathrm{eff}=4/3$, for which \citet{SchayeDallaVecchia2008} showed that both the Jeans mass and the ratio of the Jeans length to the SPH kernel are independent of the density, thus preventing spurious fragmentation due to a lack of numerical resolution. Using their pressure-dependent prescription for star formation, the observed Kennicutt-Schmidt relation, a surface density scaling law for the star formation rate that can be written as $\dot{\Sigma}_* \propto \Sigma_\mathrm{g}^n$ \citep{Kennicutt1998}, is reproduced by construction, independent of the imposed equation of state. 

The reference simulation assumes a \citet{Chabrier2003} stellar Initial Mass Function (IMF) with low and high mass cut-offs at $0.1$ and $100\,\mathrm{M}_{\sun}$, respectively. The release of hydrogen, helium and heavier elements by these stars to the surrounding gas is tracked as well: gas can be ejected through Type II SNe and stellar winds for massive stars, and Type Ia SNe and Asymptotic Giant Branch (AGB) stars for intermediate mass stars. This implementation of stellar evolution and chemical enrichment is discussed in \citet{Wiersma2009b}.

Finally, the reference simulation includes a prescription for supernova feedback, discussed in \citet{DallaVecchiaSchaye2008}. Supernovae are capable of depositing a significant amount of energy in the surrounding gas, driving large-scale winds that may eject large amounts of gas, dramatically suppressing the formation of stars. In the model used here, the energy from SNe is injected into the gas kinetically. After a delay time of $30\,\mathrm{Myr}$, a new star particle $j$ will ``kick'' a neighbouring SPH particle $i$ with a probability $\eta m_j/\sum_{i=1}^{N_\mathrm{ngb}}m_i$ in a random direction, giving it an extra velocity $v_\mathrm{w}$. The reference simulation uses the values $\eta=2$ for the initial wind mass loading and $v_\mathrm{w}=600\vunit$ for the initial wind velocity, which corresponds to $40\%$ of the available kinetic energy for our IMF.

\begin{table*}
\caption{The different variations on the reference simulation that are compared in this paper. Unless noted otherwise, all simulations use a set of cosmological parameters derived from the WMAP3 results and use identical initial conditions.}
\centering
\begin{tabular}{l c l}
\hline
Simulation &  & Description \\ [0.5ex]
\hline
\textit{AGN} &  & Includes AGN (in addition to SN feedback) \\ [1ex]
\textit{AGN\_WMAP7} &  & Same as \textit{AGN}, but with a WMAP7 cosmology \\ [1ex]
\textit{DBLIMFV1618} &  & Top-heavy IMF at high pressure, extra SN energy in wind velocity \\ [1ex]
\textit{DMONLY} &  & No baryons, cold dark matter only \\ [1ex]
\textit{DMONLY\_WMAP7} &  & Same as \textit{DMONLY}, but with a WMAP7 cosmology \\ [1ex]
\textit{MILL} &  & Millennium simulation cosmology (i.e.\ WMAP1), $\eta=4$ (twice the SN energy of \textit{REF}) \\ [1ex]
\textit{NOSN} &  & No SN energy feedback \\ [1ex]
\textit{NOSN\_NOZCOOL} &  & No SN energy feedback and cooling assumes primordial abundances \\ [1ex]
\textit{NOZCOOL} &  & Cooling assumes primordial abundances \\ [1ex]
\textit{WDENS} &  & Wind mass loading and velocity depend on gas density (SN energy as \textit{REF}) \\ [1ex]
\textit{WML1V848} &  & Wind mass loading $\eta=1$, velocity $v_\mathrm{w}=848\vunit$ (SN energy as \textit{REF}) \\ [1ex]
\textit{WML4} &  & Wind mass loading $\eta=4$ (twice the SN energy of \textit{REF}) \\ [1ex]
\hline
\end{tabular}
\vspace{10pt}
\label{simtable}
\end{table*}
\subsection{Other models}
\label{sec:othersims}
The OWLS project includes many variations on the reference simulation. We will now briefly discuss the simulations that we compare to in \S\ref{sec:barprocesses}. The different models are listed in Table~\ref{simtable}. For more details and other models we again refer to \citet{Schaye2010}.

The simulation \textit{DMONLY} includes only dark matter, hence the only active physical process is gravity. This model is useful, as many (semi-)analytical models for the matter power spectrum assume that baryons are unimportant on large scales.

The \textit{NOSN} simulation excludes supernova feedback, and the simulation \textit{NOZCOOL} assumes primordial abundances when computing cooling rates. The simulation \textit{NOSN\_NOZCOOL} excludes both SN feedback and metal-line cooling. Naturally, none of the three simulations can be considered realistic as we know that the omitted processes exist, but they are valuable tools to investigate on what scales and in what measure these processes affect the total matter power spectrum. In fact, the same may be said for the other models we consider as all, except for \textit{AGN}, suffer from the overcooling problem and hence apparently miss an important process that does occur in nature (be it AGN feedback or something else).

Supernova feedback models suffer from large uncertainties due to the limited resolution of the simulations and a lack of observational constraints. Though the product of the initial wind mass loading and the initial wind velocity squared, $\eta v_\mathrm{w}^2$, determines the energy injected into the winds per unit stellar mass and is therefore limited from above by the energy available from the SNe, the individual parameters are poorly constrained and can thus be varied. One variation on the reference model that uses the same SN energy per unit stellar mass as \textit{REF} is \textit{WML1V848}, in which the wind mass loading is reduced by a factor of $2$ while the wind velocity is increased by a factor of $\sqrt{2}$. Another such variation is \textit{WDENS}, in which the wind parameters scale with the density of the gas from which the star particle formed: the wind velocity as $v_\mathrm{w} \propto n_\mathrm{H}^{1/6}$, and the wind mass loading as $\eta \propto v_\mathrm{w}^{-2} \propto n_\mathrm{H}^{-1/3}$. Both parameters are equal to their fiducial values for stars formed at the density threshold for star formation. For the polytropic EoS that we impose onto the ISM, the wind velocity in this model scales with the local effective sound speed, as might be the case for thermally driven winds.

We also compare to models where the SN energy is varied. One scenario in which the SN energy may be higher than that in the reference model is when, under certain circumstances, the IMF becomes top-heavy, meaning that relatively more high-mass stars are produced. It is expected that the IMF is top-heavy at high redshift and low metallicity \citep[e.g.][]{Larson1998}, and both observations and theory suggest that it may be top-heavy in extreme environments like the galactic centre and starburst galaxies \citep[e.g.][]{Baugh2005, Bartko2010}. In the simulation \textit{DBLIMFV1618}, the latter effect is modelled by a switch from the Chabrier IMF to one that follows $\phi(m) \propto m^{-1}$ once the gas reaches a certain pressure threshold, which is set so that $\sim 10\%$ of the stellar mass forms with a top-heavy IMF. In this case, the emissivity in ionizing photons goes up by a factor $7.3$, and it is assumed that the SN energy scales up by the same factor. In the model we consider here, this extra energy is used to raise the wind velocity by a factor $\sqrt{7.3}$.

The final model that we consider that only differs from \textit{REF} in terms of its wind parameters is \textit{WML4}, in which the SN energy per unit stellar mass is doubled by simply increasing the wind mass loading by a factor of two. The same is done in the simulation \textit{MILL}. However, the most important feature of the latter is that it uses the same values for the cosmological parameters as the Millennium simulation \citep{Springel2005b}. These are derived from first-year WMAP data and are given by: $\{\Omega_\mathrm{m}, \Omega_\mathrm{b}, \Omega_\mathrm{\Lambda}, \sigma_8, n_\mathrm{s}, h\}$ = $\{0.25, 0.045, 0.75, 0.9, 1.0, 0.73\}$.

The last and, for our purposes, most important physics variation we consider here adds a phenomenon that has proved to be increasingly necessary to reconcile theory and observations, from the scales of individual galaxies to clusters: Active Galactic Nuclei, or AGN. They are caused by the emission of large amounts of energy from the accreting supermassive black holes that reside at the centres of galaxies, in the form of radiation that may couple to the gas and relativistic jets caused by the magnetic field of the infalling material, which can heat and displace gas out to very large distances. AGN have been invoked to explain, for example, the low star formation rates of high-mass galaxies and the suppression of cooling flows in clusters. Moreover, \citet{LevineGnedin2006} have used a toy model to demonstrate that AGN feedback may provide sufficient energy to have a large effect on the matter power spectrum.

We model the growth of supermassive black holes and the associated feedback processes using the prescription detailed in \citet{BoothSchaye2009}, which is an extension of that by \citet{Springel2005a}. During the simulation, a black hole seed particle with mass $m_\mathrm{seed}=9 \times 10^4\munit$ (i.e.\ $10^{-3}\,m_\mathrm{baryon}$) is placed at the centre of every dark matter halo whose mass exceeds $m_{\mathrm{halo},\mathrm{min}}=4 \times 10^{10}\munit$ (corresponding to $10^2$ dark matter particles). These particles then accumulate mass from the surrounding gas at an (Eddington-limited) rate based on Bondi-Hoyle-Lyttleton accretion \citep{BondiHoyle1944, HoyleLyttleton1939}, but scaled up by a factor $\alpha$ to account for the lack of a cold gas phase and the finite numerical resolution. However, for densities below our star formation threshold we do not expect a cold phase to be present and we therefore set $\alpha$ equal to unity. To ensure a smooth transition, $\alpha$ is made to depend on the density of the gas:
\begin{equation}
\alpha = \left\{\begin{array}{c l}
1 & \quad \mbox{if $n_\mathrm{H} < n_\mathrm{H}^*$}\\
\left(\frac{n_\mathrm{H}}{n_\mathrm{H}^*}\right)^\beta & \quad \mbox{otherwise.}\\
\end{array} \right.
\label{alpha}
\vspace{5pt}
\end{equation}
Here the density threshold $n_\mathrm{H}^*$ is the critical value required for the formation of a cold interstellar gas phase ($n_\mathrm{H}^*	=	0.1\,\mathrm{cm}^{-3}$; see \S \ref{sec:ref}). Models of this type are called `constant-$\beta$ models', and the fiducial value $\beta=2$ is used throughout this paper.

The black holes inject 1.5 per cent of the rest mass energy of the accreted gas into the surrounding matter in the form of heat. This feedback efficiency determines the normalisation, but not the slope, of the relations between black hole mass and galaxy properties. \citet{BoothSchaye2009} and \citet{BoothSchaye2011} demonstrate that this efficiency reproduces the observed relations between BH mass and both stellar mass and stellar velocity dispersion, as well as their evolution. \citet{McCarthy2010} have shown that the \textit{AGN} simulation, but not the reference model, provides excellent agreement with both optical and X-ray observables of groups of galaxies at redshift zero. In particular, it reproduces the temperature, entropy, and metallicity profiles of the gas, the stellar masses, star formation rates, and age distributions of the central galaxies, and the relations between X-ray luminosity and both temperature and mass. We therefore consider simulation \textit{AGN} to be more realistic than our other models. As we shall see in \S\ref{sec:results}, the inclusion of AGN feedback greatly affects the power spectrum on a large range of scales.

Finally, we have re-run two simulations, \textit{DMONLY} and \textit{AGN}, with cosmological parameters derived from the WMAP 7-year results	\citep{Komatsu2011}: \{$\Omega_\mathrm{m}$, $\Omega_\mathrm{b}$, $\Omega_\mathrm{\Lambda}$, $\sigma_8$, $n_\mathrm{s}$, $h$\} = \{$0.272$, $0.0455$, $0.728$, $0.81$, $0.967$, $0.704$\}. These versions are called \textit{DMONLY\_WMAP7} and \textit{AGN\_WMAP7}, respectively. We consider the latter to be our most realistic and up-to-date model. Note that the linear input power spectra used for the initial conditions of these simulations have not been generated by \textsc{cmbfast}, but by the more up-to-date \textsc{f90} package \textsc{camb} \citep[][, version January 2010]{Lewis2002}.

\subsection{Power spectrum calculation}
The distribution of matter in the Universe can be described by a continuous density function, $\rho(\mathbf{x})$, where the vector $\mathbf{x}$ specifies the position relative to some arbitrary origin. Given this density field, we consider fluctuations, $\delta(\mathbf{x})$, defined as:
\begin{equation}
\delta(\mathbf{x})\equiv\frac{\rho(\mathbf{x})-\bar{\rho}}{\bar{\rho}},
\label{fluctuation}
\end{equation}
where $\bar{\rho}$ is the global mean density. We can relate this density contrast to wave modes $\hat{\delta}_\mathbf{k}$ via a discrete Fourier transform:
\begin{equation}
\delta(\mathbf{x})=\sum_\mathbf{k}\hat{\delta}_\mathbf{k}\,e^{-i\mathbf{k}\cdot\mathbf{x}}.
\label{Fourier1}
\end{equation}
The density field can thus be seen as made up of waves with certain amplitudes and phases, with wave vectors $\mathbf{k}$. We now define the power spectrum, $P(k)$, as:
\begin{equation}
P(k) \equiv V \left<|\hat{\delta}_\mathbf{k}|^2\right>_\mathrm{k},
\label{Pk}
\end{equation}
where $V$ is the volume under consideration. The power spectrum is therefore obtained by collecting the amplitudes-squared of all wave modes with the same wave number $k=|\mathbf{k}|$, and averaging them. This makes it clear that the power spectrum is a statistical tool, whose accuracy increases when more waves of the same length are available (i.e.\ when the scale $2\pi/k$ is small compared to the size of the box). We will present our results using what is often called the dimensionless matter power spectrum, which is defined as:
\begin{equation}
\Delta^2(k)=\frac{k^3}{2\pi^2}P(k).
\label{convergence}
\end{equation}
The dimensionless power spectrum scales with the mass variance, $\sigma^2(M)$, where $M=\frac{4\pi}{3}\left(\frac{2\pi}{k}\right)^3\bar{\rho}$. Note that using $\Delta^2(k)$ instead of $P(k)$ does not affect the relative differences between power spectra.

The code we have chosen to use to obtain accurate power spectrum estimations from our simulations is the publicly available \textsc{f90} package called \textsc{powmes} \citep{Colombi2009}. The advantages of \textsc{powmes} stem from the use of the Fourier-Taylor transform, which allows analytical control of the biases introduced, and the use of foldings of the particle distribution, which allow the dynamic range to be extended to arbitrarily high wave numbers while keeping the statistical errors bounded. For a full description of these methods we refer to \citet{Colombi2009}. We have compared the performance of \textsc{powmes} with respect to power spectrum estimators using simple NGP, CIC and TSC interpolation schemes, and found that \textsc{powmes} is capable of obtaining far more accurate power spectra over a larger spectral range within the same computation time.

We have expanded \textsc{powmes} with the possibility to consider only one group of particles at a time, in order to see which parts of the power spectrum are dominated by the contribution of, for example, cold dark matter (see \S\ref{sec:contributions}). Finally, we performed extensive timing tests using different grids, foldings, CPUs and particle numbers which, combined with the performance results from \citet{Colombi2009}, resulted in the fiducial values $G=256$ and $F=7$ for the number of grid points on a side and the number of foldings, respectively.

\subsubsection{Discreteness and other numerical limitations}
In Appendix~A we demonstrate that the simulations are sufficiently converged with respect to increases in the numerical resolution to predict the power spectrum with better than 1\% accuracy for $k \la 10\kunit$. This range is greatly expanded in both directions if we only consider the relative differences in power between simulations.

Besides numerical resolution, the predicted power spectra may be affected by sample variance, which is generally called cosmic variance in cosmology. This is caused by the finite volume of the box and by the fact that each simulation provides only a single realisation of the underlying statistical distribution. Note that finite volume effects are different for the simulations than for observational surveys, because the mean density in the simulation boxes is always equal to the cosmic mean. In Appendix~A we show that finite volume effects may cause us to  underestimate the effects of baryons on scales of several tens of Mpc, i.e.\ close to the size of the box. While the fact that we only use a single realisation of the initial conditions prevents us from obtaining highly accurate \textit{absolute} values for the power spectrum on scales close to the box size, it does not prevent us from investigating the \textit{relative} changes in power caused by baryon physics.

Finally, we are limited in our determination of the power spectrum by the discreteness of the density field. Because we use particles to represent a continuous field, there will always be non-zero power present at all scales, called white noise or shot noise. If we assume the particle distribution to be a local Poisson realisation of a stationary random field, an assumption used in any calculation of the power spectrum and one that is expected to be valid for an evolved distribution,\footnote{The discreteness noise can initially be much smaller if the particles are arranged on a grid or in a ``glass-like'' fashion. Particles in low-density regions may retain memory of their initial distribution, reducing the noise below the level expected for a Poisson distribution.} this white noise component can be calculated \citep[see, for example,][]{Peebles1980,Peebles1993,Colombi2009}. Subtracting the shot noise from the initial estimate of the power spectrum will make the latter somewhat more accurate, but one should still expect the uncertainty on the estimate of the power spectrum to increase dramatically when the intrinsic power spectrum falls far below the shot noise level. The contribution of shot noise to $P(k)$ is independent of $k$. Hence, if we use $\Delta^2(k)$ as the measure of the power spectrum, then the shot noise level will scale as $k^3$. In the following section, the scale at which the shot noise of each simulation is equal to (the white noise corrected) $\Delta^2(k)$ is denoted by a circle, while it is shown explicitly in Appendix~A. Note that the theoretical shot noise level has been subtracted for all power spectra shown in this paper.

\section{Results}
\label{sec:results}
In this section we present the power spectra obtained from our simulations. In \S\ref{sec:dmonly} we compare the power spectrum of our dark matter only simulation to predictions from the literature. We investigate the effects of adding or modifying prescriptions for baryonic processes in \S\ref{sec:barprocesses}. We examine how well CDM, gas, and stars trace each other and consider the contributions of these different components to the total power in the reference simulation in \S\ref{sec:contributions}, and we examine the back-reaction of baryons on the CDM for the two most important simulations, \textit{REF} and \textit{AGN}, in \S\ref{sec:backreaction}. Finally, in \S\ref{sec:agn}, we take a closer look at model \textit{AGN}, which we consider to be our most realistic simulation because it reproduces the optical and X-ray observations of groups of galaxies \citep{McCarthy2010}. We investigate the effect of using the WMAP7 rather than the WMAP3 cosmology, compare to widely used model power spectra, and consider the evolution of the effect of baryons on the matter power spectrum.

\begin{figure}
\begin{center}
\includegraphics[width=0.95\columnwidth, trim=0mm -15mm 0mm -20mm]{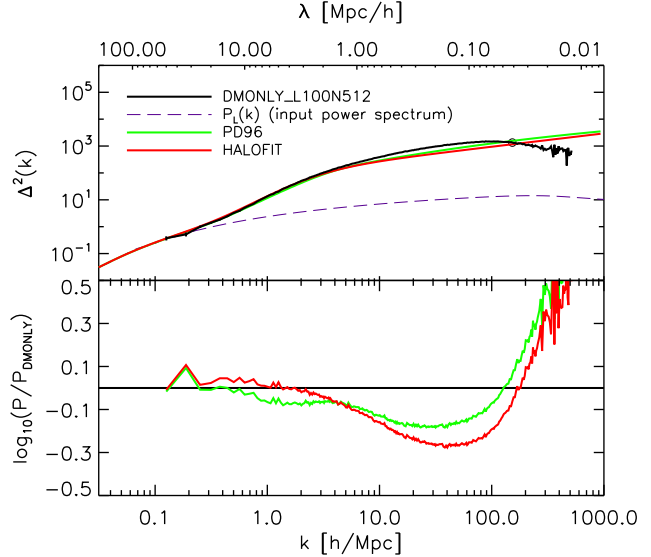}
\caption{Comparison of the matter power spectrum of \textit{DMONLY\_L100N512} with analytical fits by \citet[][, PD96]{PeacockDodds1996} and \citet[][, HALOFIT]{SmithPeacock2003} at redshift zero. The small circle, drawn in this and all following plots showing $\Delta^2(k)$, indicates the scale below which the (subtracted) shot noise in the simulation becomes significant, and the dashed purple curve shows the linear input power spectrum of the simulations. The bottom panel shows the ratios of the power spectra from theoretical models and the simulation. There is good agreement down to scales of a few Mpc, especially for the more recent HALOFIT model, but on smaller scales \textit{DMONLY} predicts up to twice as much power as HALOFIT. For $\lambda < 10^2\lunitk$ the power in the \textit{DMONLY} simulation drops due to a lack of resolution.}
\label{models1}
\end{center}
\end{figure}

\subsection{Comparison of a dark matter only simulation to models}
\label{sec:dmonly}
In this section we compare the power spectrum of our \textit{DMONLY} simulation to those predicted by the widely used models of \citet[][, hereafter PD96]{PeacockDodds1996} and \citet[][, hereafter HALOFIT]{SmithPeacock2003}.

The PD96 model is an extension of what is known as the HKLM model \citep{Hamilton1991}, which first introduced a universal analytical formula to map the linear correlation function into a non-linear one, the coefficients of which were estimated using N-body simulations. Both of these models assume spherical collapse of fluctuations that have reached a certain overdensity, followed by stable clustering (which states that the mean physical separation of particles is constant on sufficiently small scales). \citet{PeacockDodds1994}, followed by PD96, expanded on the groundwork laid by HKLM by presenting a version of the method that worked with power spectra instead and allowed for $\Omega \neq 1$, a non-zero cosmological constant and large negative spectral indices.

However, numerical simulations have shown that the assumption of stable clustering is not always valid. The more recent HALOFIT model by \citet{SmithPeacock2003} aimed to improve on PD96 by taking this into account. This method is based on concepts from the ``halo model'', in which the density field is viewed as a distribution of isolated haloes \citep[e.g.][]{PeacockSmith2000, Seljak2000, CooraySheth2002}. It is then assumed that the power spectrum can be split into two parts: a large-scale quasi-linear term that is due to the clustering of separate haloes (the 2-halo term), and a small-scale term caused by the correlation of subhaloes within the same parent halo (the 1-halo term). Their resulting analytical formulae were fit to power spectra obtained from N-body simulations.

To create power spectra that conform with these models and the cosmological parameters used in our simulations, we have utilised the publicly available package \textsc{iCosmo}, described in \citet{Refregier2008}. We chose to generate the linear power spectra using the \citet[][, EH]{EisensteinHu1999} transfer function. We have also tried using the \citet[][, BBKS]{Bardeen1986} transfer function to generate initial conditions for the PD96 model, as this is the one originally used by the authors, which introduced only minor differences with respect to the results shown here ($1-10\%$ lower power for $k \ga 10\kunit$).

In Figure~\ref{models1} we compare these models to the simulation that, like the theoretical models for non-linear growth, includes only dark matter (\textit{DMONLY}). For reference, the dashed curve shows the linear input power spectrum of the simulations. The bottom panel shows the ratio of the analytical predictions to our results. Note that we have omitted the first wave mode (at $\lambda = 100\lunit$) in all of our figures because we cannot sample the power spectrum on the scale of the simulation box. We see that the dark matter power spectrum follows the analytical predictions pretty well on large scales (except on the scale of the simulation box), and that HALOFIT provides a better match than the PD96 model, as expected. However, on scales below a few Mpc the theoretical models start to severely underestimate the amount of structure formed in the simulation, and the difference between HALOFIT and the \textit{DMONLY} simulation reaches a factor of 2 on scales of $1-3 \times 10^{-1}\lunit$. The rapid decline of the \textit{DMONLY} power spectrum for $k \ga 100\kunit$ ($\lambda < 10^2\lunitk$) is mostly due to the underproduction of low-mass haloes due to the finite resolution (see Appendix~A). While we will always show the power spectrum up to $k\approx 500\kunit$, we are mainly interested in the scales relevant for upcoming surveys, $k \la 10\kunit$. As discussed in Appendix~A, for $k\gg 10\kunit$ numerical convergence may become an issue. Note that the power spectrum of the simulation remains reasonably well-behaved far below the theoretical shot noise level (i.e.\ well to the right of the small circle), indicating that the subtraction of this noise component is fairly accurate.

Newer implementations of the halo model exist, based on fits to more recent N-body simulations. These models improve on the HALOFIT model by including a variable concentration-mass relation \citep[such as those derived by][]{Neto2007, Duffy2008} and have been shown to reproduce the power spectra from simulations with higher accuracy \citep[e.g.][]{Hilbert2009}. Since no suitable codes employing these models were available, we do not compare to their results here. However, as \citet{Hilbert2009} have shown that using the halo model with the concentration-mass relation of \citet{Neto2007} increases the power at intermediate scales, we suspect that such models would provide a better match to the power spectrum of \textit{DMONLY}.

\begin{figure}
\begin{center}
\includegraphics[width=0.95\columnwidth, trim=7mm -17mm 0mm -20mm]{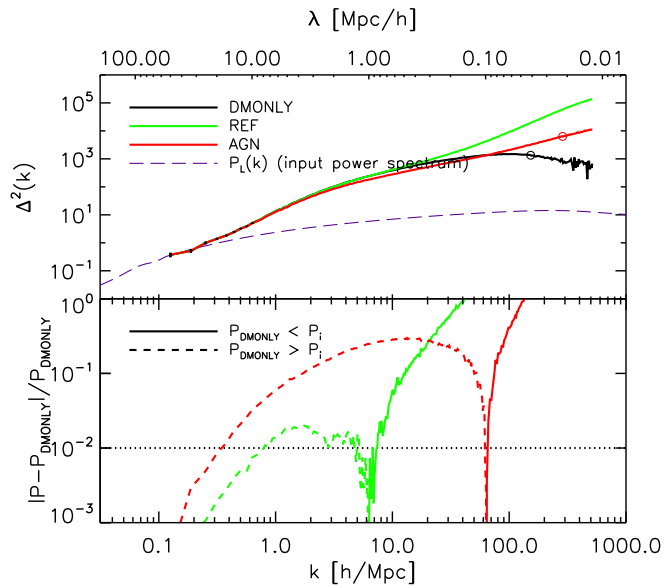}
\caption{A comparison of the total matter power spectra of \textit{DMONLY\_L100N512} (black), \textit{REF\_L100N512} (green) and \textit{AGN\_L100N512} (red), at redshift $z=0$. The bottom panel shows the absolute value of the relative difference of the latter two with respect to \textit{DMONLY}; solid (dashed) curves indicate that the power is higher (lower) than for \textit{DMONLY}. The dotted, horizontal line shows the $1\%$ level. Note that the first wave mode has been omitted as it holds no information. While pressure forces smooth the baryonic density field on intermediate scales, cooling allows the baryons to increase the total power on small scales. The addition of AGN feedback, which is required to match observations of groups, has an enormous effect, reducing the power by $\ga 10\%$ for $k\ga 1\kunit$.}
\label{DMONLYREFAGN}
\end{center}
\end{figure}

\subsection{The relative effects of different baryonic processes}
\label{sec:barprocesses}
In this section we present our main results, demonstrating how single baryonic processes, or implementations thereof, can influence the matter power spectrum. While we will focus mainly on the range of scales relevant to upcoming weak lensing surveys, $0.1\kunit < k < 10\kunit$, we will also discuss the differences at the much smaller scales that our simulations allow us to probe. We note again that all power spectra are taken from simulations with $L=100\lunit$ and $N=512$ at redshift zero, and that, unless stated otherwise, all simulations are evolved from the same initial conditions.

We start by comparing the power spectra of \textit{DMONLY}, the reference simulation (\textit{REF}) and \textit{AGN}, in Figure~\ref{DMONLYREFAGN}. The panel at the bottom of most plots in this section shows the absolute value of the relative difference between power spectra. The dotted, horizontal line shows the $1\%$ level: any differences above this level will thus affect the statistics of surveys that aim to measure the power spectrum to this accuracy.

It is immediately clear from the comparison between \textit{DMONLY} and \textit{REF} that the contribution of the baryons is significant, decreasing the power by more than $1\%$ for $k \approx 0.8-5\kunit$. This is because gas pressure smooths the density field relative to that expected from dark matter alone. On scales smaller than $1\lunit$ ($k \ga 6\kunit$), the power in the reference simulation quickly rises far above that of the dark matter only simulation, because radiative cooling enables gas to cluster on smaller scales than the dark matter. These results confirm the findings of previous studies, at least qualitatively \citep[e.g.][]{Jing2006, Rudd2008, Guillet2009, Casarini2010}.

However, when AGN feedback is included, the results change drastically. In this case, the reduction in power relative to \textit{DMONLY} already reaches $1\%$ for $k \approx 0.3\kunit$ ($\lambda \approx 20\lunit$) and exceeds $10\%$ for $2 \la k \la 50\kunit$. We thus see that AGN feedback even suppresses the total matter power spectrum on very large scales. The enormous effect of AGN feedback is due to the removal of gas from (groups of) galaxies. That large amounts of gas are indeed being moved to large radii in this simulation has been shown by, for example, \citet[][, e.g.\ Figures 1 \& 2]{Duffy2010} and \citet[][, e.g.\ Figure 3]{McCarthy2011}. Because the AGN reside in massive and thus strongly clustered objects, the power is suppressed out to scales that exceed the scale on which individual objects move the gas. 

\begin{figure}
\begin{center}
\includegraphics[height=0.6\textheight, trim=50mm -8mm 55mm -8mm]{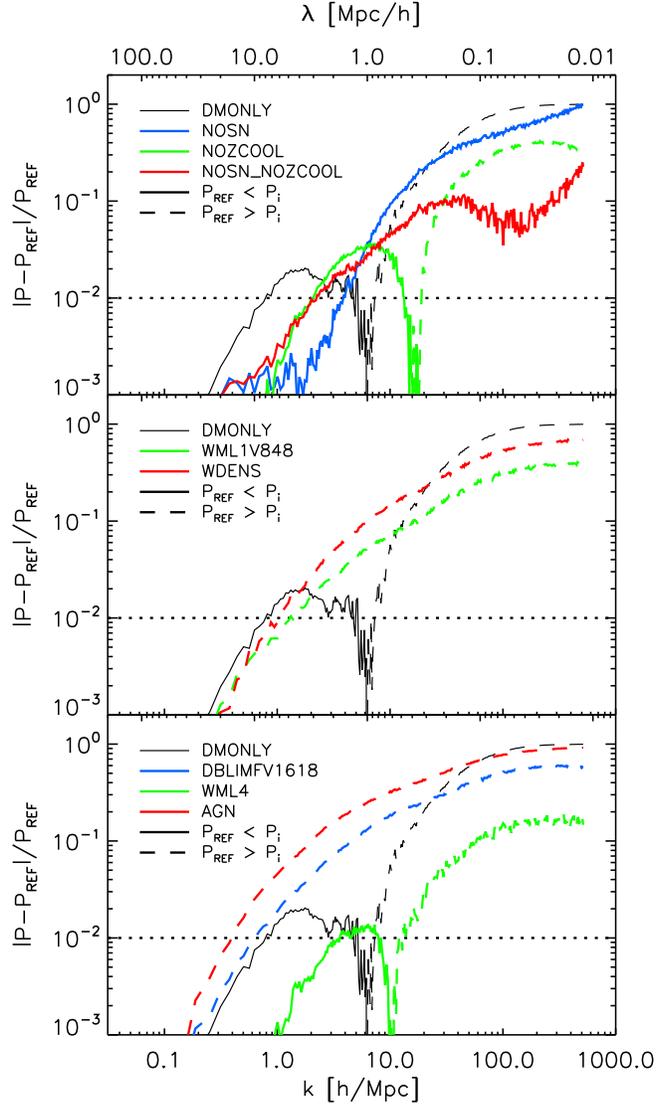}
\caption{Comparisons of $z=0$ power spectra predicted by simulations incorporating different physical processes to that predicted by the reference simulation. The panels are similar to the bottom panel of Figure~\ref{DMONLYREFAGN}, but now show differences relative to \textit{REF}. The thin black curve that is repeated in all panels shows the relative difference with \textit{DMONLY}. Colours indicate different simulations, while different line styles indicate whether the power is reduced or increased relative to the reference simulation. \newline
\textit{Top:} A simulation without SN feedback (blue), one without metal-line cooling (green) and one that excludes both effects (red). SN feedback decreases the power on all scales. Metal-line cooling decreases the power for $\lambda > 0.4\lunit$ but increases the power on smaller scales. The effects of removing both SN feedback and metal-line cooling are $> 10\%$ for $k>20\kunit$ and $>1\%$ for $k>2\kunit$. \newline
\textit{Middle:} Different SN wind models which all use the same amount of SN energy per unit stellar mass (see text). The effects of varying the implementation of SN feedback, while keeping the SN energy that is injected per unit stellar mass the same, are $>10\%$ for $k>10\kunit$ and $>1\%$ for $k>1\kunit$. \newline
\textit{Bottom:} Models with different feedback energies and processes, see text for details. Including a top-heavy IMF at high pressure (\textit{DBLIMFV1618}) or AGN feedback (\textit{AGN}) greatly reduces the power. The reduction caused by the latter is $>10\%$ for $k>2\kunit$ and $>1\%$ for $k>0.4\kunit$.}
\label{REF_compare}
\end{center}
\end{figure}

Figure~\ref{REF_compare} shows the difference in the power spectra predicted by a variety of simulations relative to that predicted by the reference simulation. The models are listed in Table~\ref{simtable} and were described in \S\ref{sec:othersims}. The top panel shows the effect of turning off SN feedback and/or metal-line cooling. Since SN feedback heats and ejects gas, we expect it to decrease the small-scale power. Indeed, the power in \textit{NOSN} is $> 1\%$ higher than in the reference simulation for $k > 4\kunit$ and the difference reaches $10\%$ at $k \approx 10\kunit$. The absence of SN feedback also increases the star formation rate, making stars the dominant contributor to the total matter power spectrum out to larger scales (not shown). 

Turning off metal-line cooling reduces the power on small scales because less gas is able to cool down and accrete onto galaxies. Indeed, model \textit{NOZCOOL} predicts $10-50\%$ less power for $k \ga 30\kunit$. However, the absence of metal-line cooling increases the power by several percent for $\lambda \sim 1\lunit$ because the lower cooling rates force more gas to remain at large distances from the halo centres. 

Even though the effects of SN feedback and metal-line cooling are somewhat opposite in nature, as the former increases the energy of the gas while the latter allows the gas to radiate more of its thermal energy away, removing both processes in the simulation \textit{NOSN\_NOZCOOL} still introduces differences of about $1-10\%$ for $k \ga 2\kunit$ relative to the reference simulation. It is therefore vital to take both SN feedback and metal-line cooling into account if one wants to predict the matter power spectrum with an accuracy better than 10\%.

We compare models that use different prescriptions for SN feedback, but the same amount of SN energy per unit stellar mass as \textit{REF}, in the middle panel of Figure~\ref{REF_compare}. In \textit{WML1V848} the SN energy is distributed over half as much gas, but the initial wind velocity is a factor $\sqrt{2}$ higher, resulting in more effective SN feedback in all but the lowest mass galaxies. The differences with respect to the reference model extend to even larger scales than when SN feedback is removed entirely: the power is reduced by $>1\%$ for $k \ga 1\kunit$ and by $\ga 10\%$ for $k \ga 10\kunit$. In model \textit{WDENS} the initial wind velocity increases with the local sound speed in the ISM, but the mass loading is adjusted so as to keep the amount of SN energy per unit stellar mass equal to that in \textit{REF}. This implementation results in an even stronger decrease in power on scales $< 10\lunit$. In both these models, the reduction in power is caused by the increased effectiveness of SN feedback in driving outflows of gas. We stress that because of our lack of understanding of the effects of SN feedback, there is a priori no reason to assume that the model used in the reference simulation is a better approximation to reality than the models we compare to here.

\begin{figure}
\begin{center}
\includegraphics[width=0.95\columnwidth, trim=7mm 15mm 5mm -20mm]{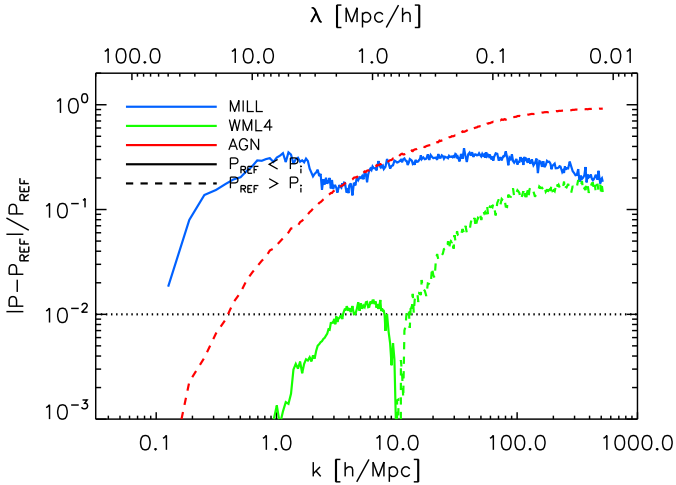}
\caption{Difference of the $z=0$ matter power spectrum in a simulation using a WMAP1 cosmology (\textit{MILL}) relative to that of the \textit{REF} model, which assumes the WMAP3 cosmology, after rescaling the former to match the latter on the scale of the simulation box ($\lambda=100\lunit$, not shown). \textit{WML4} is shown for reference as this simulation uses the same baryonic physics as \textit{MILL}. For $k \ga 3\kunit$, the effect of AGN feedback is at least as strong as that of this unrealistically large change in cosmology.}
\label{MILL}
\end{center}
\end{figure}

In fact, it is possible that the SN energy per unit stellar mass is different from the value assumed in the \textit{REF} model, or that it varies with environment. Model \textit{DBLIMFV1618}, which we compare with \textit{REF} in the bottom panel of
Figure~\ref{REF_compare}, uses a top-heavy IMF in high-pressure environments. Such an IMF yields more SNe per unit stellar mass which decreases the power by $> 1$\% for $k>0.7\kunit$ and by $>10\%$ for $k>4\kunit$. Clearly, it will be necessary to understand any environmental dependence of the IMF in order to predict the matter power spectrum to $1\%$ accuracy on the scales relevant for upcoming surveys.

On the other hand, doubling the wind mass loading, while keeping the wind velocity fixed to the value used in \textit{REF}, as is done in \textit{WML4}, has a far more modest effect. This is because the wind velocities are too low to significantly disturb the high-pressure ISM of massive galaxies. The differences with respect to the reference model are limited to $\la 1\%$ for $k \la 10\kunit$.

The bottom panel of Figure~\ref{REF_compare} also compares the reference simulation to model \textit{AGN}, which differs from \textit{REF} by the inclusion of a phenomenon that has been shown to play a role in many contexts and that strongly improves the agreement with observations of groups of galaxies \citep{McCarthy2010}. Like SN feedback, AGN feedback decreases the power by heating and ejecting gas, but the effect is more dramatic than that of the standard SN feedback model, both in scope and magnitude. With respect to the reference model, the power is decreased by $\ga 30\%$ for $k>10\kunit$ and by $\ga 5\%$ for $k>1\kunit$. The reduction in power only falls below $1\%$ for $k < 0.4\kunit$ ($\lambda \ga 10\lunit$). Note that the effect of AGN feedback is strikingly similar to, albeit stronger than, that of the stellar feedback model that uses a top-heavy IMF in high-pressure environments.

It is clear that many different baryonic processes, and even slightly different implementations thereof, are capable of introducing significant differences in the matter power spectrum on scales relevant for observational cosmology. To put the effects of baryons into perspective, we compare to a simulation with a very different cosmology, \textit{MILL}, in Figure~\ref{MILL}. The difference between the cosmology derived from the first-year WMAP data used in \textit{MILL} and the one derived from the 3-year WMAP data used in the other simulations is large; in fact, the difference is much larger than the error bars of the most recent data allow. For reference, we note that the currently favoured cosmology \citep{Komatsu2011} lies in between those given by WMAP1 and WMAP3. To account for the difference in normalisation of the \textit{MILL} power spectrum, which is caused mainly by its higher $\Omega_\mathrm{m}$ and $\sigma_8$ values, we have rescaled it to have the same power at the box size as \textit{REF}. Still, the effect on the power spectrum exceeds $10\%$ for $k \ga 0.2\kunit$. A quick comparison with \textit{WML4}, which uses the exact same baryon physics as \textit{MILL} and twice the SN wind mass loading used in \textit{REF}, shows that the effect of the change in mass loading is relatively small, as we had already shown in Figure~\ref{REF_compare}. However, we see that for $k \ga 3\kunit$, the effect of AGN feedback is at least as strong as that of this unrealistically large change in cosmology. We thus conclude that baryonic effects are not only significant at the $\sim 1\%$ level, but can even be larger than a ``very wrong'' choice of cosmology.

Almost all theoretical models used in the literature consider only CDM, assuming that the baryons follow the dark matter perfectly for $k \la 1\kunit$. We have shown (see Fig.~\ref{DMONLYREFAGN}) that the fact that baryons experience gas pressure reduces the power on large scales, while their ability to radiate away their thermal energy increases the power on small scales. If we ignore AGN feedback, as has been done in all previous work, we find that the power is reduced by at least a few percent for $0.8 < k <  5\kunit$ and that the power is increased for $k > 7\kunit$, with the difference reaching approximately $6\%$ at $k=10\kunit$ for the reference model. However, the single process of AGN feedback, which improves the agreement with observations of groups of galaxies, reduces the power by $\ga 10\%$ over the whole range $1 \la k \la 10\kunit$ and the reduction only drops below 1\% for $k< 0.3\kunit$. Highly efficient SN feedback, as may for example result from a top-heavy IMF in starbursts, would have nearly as large an effect. One can therefore not expect to constrain the primordial power spectrum more accurately until such processes are better understood and included in theoretical models.

\begin{figure*}
\begin{center}
\includegraphics[width=0.85\textwidth, trim=13mm 34mm 23mm -10mm]{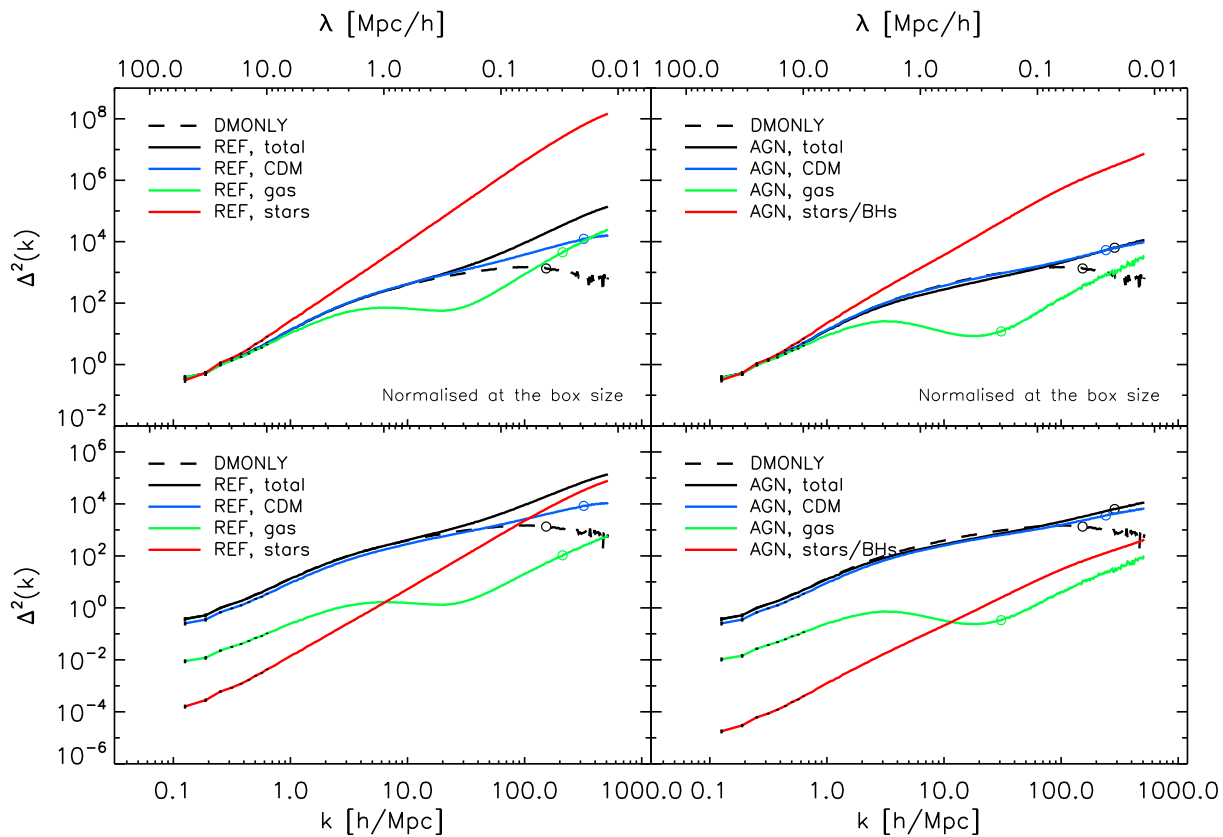}
\caption{Decomposing the $z=0$ total power spectra (black) into the contributions from cold dark matter (blue), gas (green) and stars/black holes (red). The left and right columns show results for \textit{REF\_L100N512} and \textit{AGN\_L100N512}. In the top row the density contrast of each component $i$ is defined relative to its own mean density, i.e.\ $\delta_i \equiv (\rho_i - \bar{\rho}_i)/\bar{\rho}_i$. This guarantees that all power spectra converge on large scales, thus enabling a straightforward comparison of their shapes. In the bottom row the density contrast of each component is defined relative to the total mean density, i.e.\ $\delta_i \equiv (\rho_i - \bar{\rho}_\mathrm{tot})/\bar{\rho}_\mathrm{tot}$, which allows one to compare their contributions to the total power. The power spectrum of the gas flattens or even decreases for $\lambda\la 1\lunit$ as a result of pressure smoothing, but its ability to cool allows it to increase again on galaxy scales ($\lambda \la 10^2\lunitk$). The power spectrum of the stellar component, which is a product of the collapse of cooling gas, increases most rapidly towards smaller scales. While stars dominate the total power for $\lambda \ll 10^2\lunitk$ in \textit{REF}, dark matter dominates on all scales when AGN feedback is included.}
\label{REFAGN_ALL}
\end{center}
\end{figure*}

\begin{figure}
\begin{center}
\includegraphics[width=0.95\columnwidth, trim=13mm 35mm 87mm -10mm]{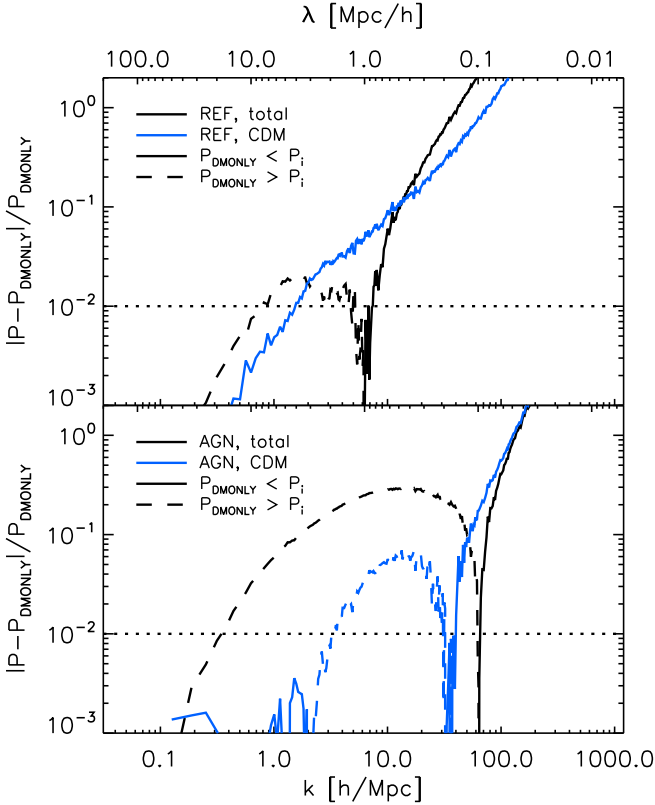}
\caption{The back-reaction of baryons on the CDM. The blue curves show the relative difference between the power spectrum of the CDM component, after scaling the CDM density by the factor $\Omega_\mathrm{m}/(\Omega_\mathrm{m}-\Omega_\mathrm{b})$, and that of a dark matter only simulation for either the \textit{REF} (top panel) or \textit{AGN} (bottom panel) model. For comparison, the relative differences between the total matter power spectra of the baryonic simulations and \textit{DMONLY} is shown by the black curves. Baryons increase the small-scale power in the CDM component. However, when AGN feedback is included, the power in the CDM component drops $1-10\%$ below that of the \textit{DMONLY} simulations for $0.2 \lunit \la \lambda \la 2\lunit$.}
\label{REFAGN_BACK}
\end{center}
\end{figure}

\subsection{Contributions of dark matter, gas and stars}
\label{sec:contributions}
Generally, power spectra are calculated using all matter inside the computational volume. This total matter power spectrum is what is measurable using e.g.\ gravitational lensing surveys. However, as we have a larger freedom of measurement using simulations, we can also consider the power in different components, for example to see which parts of the power spectrum are dominated by baryonic matter or how baryons change the distribution of cold dark matter.

On sufficiently large scales the baryons will trace the dark matter. Hence, when averaged over these scales, the baryonic and CDM densities are given by 
\begin{eqnarray}
\nonumber
\rho_\mathrm{cdm} &=& \frac{\Omega_\mathrm{m}-\Omega_\mathrm{b}}{\Omega_\mathrm{m}}\rho_\mathrm{tot},\\
\rho_\mathrm{bar} &=& \frac{\Omega_\mathrm{b}}{\Omega_\mathrm{m}}\rho_\mathrm{tot}.
\label{dens_comp}
\end{eqnarray}
We can now use these expressions to estimate the relative contributions of correlations between particle types to the total matter power spectrum. Using 
$P_\mathrm{tot}(k) \propto \left <\left |\hat{\rho}_\mathrm{tot}(k)\right |^2\right > 
\propto \left <\left |\hat{\rho}_\mathrm{cdm}\right |^2\right > + \left <\hat{\rho}_\mathrm{cdm}\hat{\rho}^\ast_\mathrm{bar}\right > + \left
<\hat{\rho}_\mathrm{cdm}^\ast\hat{\rho}_\mathrm{bar}\right >+ \left <\left |\hat{\rho}_\mathrm{bar}\right |^2\right >$, we find, for sufficiently small $k$:
\begin{eqnarray}
\nonumber
P_\mathrm{cc} &=& \frac{(\Omega_\mathrm{m}-\Omega_\mathrm{b})^2}{\Omega_\mathrm{m}^2}P_\mathrm{tot} \approx 0.68 P_\mathrm{tot},\\
P_\mathrm{cb}+P_\mathrm{bc} &=& \frac{2\Omega_\mathrm{b}(\Omega_\mathrm{m}-\Omega_\mathrm{b})}{\Omega_\mathrm{m}^2}P_\mathrm{tot} \approx 0.29 P_\mathrm{tot},\\
\nonumber
P_\mathrm{bb} &=& \frac{\Omega_\mathrm{b}^2}{\Omega_\mathrm{m}^2}P_\mathrm{tot} \approx 0.03 P_\mathrm{tot}.
\label{cross}
\end{eqnarray}
Hence, on large scales we expect the power due to the auto-correlation of CDM to dominate the total matter power spectrum, with a significant contribution from the cross terms $P_\mathrm{cb}$ and $P_\mathrm{bc}$.

The four panels of Figure~\ref{REFAGN_ALL} show power spectra for the \textit{REF\_L100N512} (left) and \textit{AGN\_L100N512} (right) simulation at $z=0$, both for the total matter (solid black) and for individual components (coloured curves). For reference, we also show the power spectrum for \textit{DMONLY\_L100N512} (dashed black). The top row shows the power spectra of $\delta_i \equiv (\rho_i - \bar{\rho}_i)/\bar{\rho}_i$. This definition ensures that the power spectra of all components $i$ converge on large scales, which allows us to examine how well different components trace each other. The bottom row, on the other hand, shows the power spectra of $\delta'_i \equiv (\rho_i-\bar{\rho}_\mathrm{tot})/\bar{\rho}_\mathrm{tot}$, which allows us to estimate the contributions of different components to the total matter power spectrum.

Looking at the top-left panel, we see that, as expected, the baryonic components trace the dark matter well at the largest scales. However, significant differences exist for $\lambda \la 10\lunit$. Observe that, at scales of several hundred kpc and smaller, the difference between \textit{REF} and the dark matter only simulation is larger than that between the latter and the analytical models we compared to earlier (see Fig.~\ref{models1}). In fact, the difference between the cold dark matter component of the reference simulation and \textit{DMONLY} is also larger than that between the latter and the analytic models. This is due to the back-reaction of the baryons on the dark matter, which we will discuss in \S\ref{sec:backreaction}.

Next, we turn to the bottom-left panel of Figure~\ref{REFAGN_ALL} which shows that cold dark matter dominates the power spectrum on large scales, as expected, although the contribution from the CDM-baryon cross power spectrum (not shown) is important as well. The contribution of baryons is significant for $\lambda \la 10^2\lunitk$ and dominates below $60\lunitk$. The strong small-scale baryonic clustering is the direct consequence of gas cooling and galaxy formation. Taking a look at how the baryonic component is itself built up, we see that gas dominates the baryonic power spectrum on large scales, but that stars take over for $\lambda < 1\lunit$. The gas power spectrum flattens for $\lambda \la 1\lunit$, which corresponds to the virial radii of groups of galaxies, but steepens again for $\lambda \la 0.1\lunit$, i.e.\ galaxy scales. The reason for the decrease in slope around $1\lunit$ is threefold. First, the pressure of the hot gas smooths its distribution on the scales of groups and clusters of galaxies. Second, as the gas collapses it fragments and forms stars. Third, due to stellar feedback the gas is distributed out to large distances, reducing the power. 

The inclusion of AGN feedback greatly impacts the matter power spectrum on a wide range of scales. Comparing the top panels of Figure~\ref{REFAGN_ALL}, we see that AGN feedback strongly decreases the power in the gas and stellar components relative to that of the dark matter for $\lambda \la 1\lunit$. A comparison of the bottom panels reveals that the contribution of stars to the total power is reduced the most, with the reduction factor increasing from an order of magnitude on the largest scales to more than two orders of magnitude on the smallest scales. This clearly shows that AGN feedback suppresses star formation, as required to solve the overcooling problem. For the gas component the change is also dramatic. While $\Delta_\mathrm{gas}^2(k)=1$ for $\lambda \sim 3\lunit$ in \textit{REF}, this level of gas power is only reached at $100\lunitk$ for \textit{AGN}. The suppression of baryonic structure by AGN feedback makes dark matter the dominant component of the power spectrum on all scales shown, although it is important to note that the dark matter distribution is also significantly affected by the AGN, as we shall see next.

\subsection{The back-reaction of baryons on the dark matter}
\label{sec:backreaction}
Even though dark matter is unable to cool through the emission of radiation, its distribution can still be altered by the inclusion of baryons due to changes in the gravitational potential. We examine this back-reaction of the baryons on the dark matter for the reference and AGN simulations in the left and right panels of Figure~\ref{REFAGN_BACK}, respectively. In order to make a direct comparison, we have rescaled the density of the dark matter component of the simulations that include baryons by multiplying it by the factor $\Omega_\mathrm{m}/(\Omega_\mathrm{m}-\Omega_\mathrm{b})$. The blue curve shows the relative differences between the power spectrum of the rescaled CDM component and that of \textit{DMONLY}.

On scales $k \ga 2\kunit$, corresponding to spatial scales $\lambda \la 3\lunit$, the power in CDM structures in the reference simulation is increased by $>1\%$ with respect to \textit{DMONLY}. The difference continues to rise towards higher $k$, reaching $10\%$ around $k=10\kunit$. Because the baryons can cool, they are able to collapse to very high densities, and in the process they steepen the potential wells of virialized dark matter haloes, causing these to contract. The effect is larger closer to the centres of these haloes, i.e.\ on smaller scales.

\begin{figure}
\begin{center}
\includegraphics[width=0.95\columnwidth, trim=7mm 15mm 5mm -20mm]{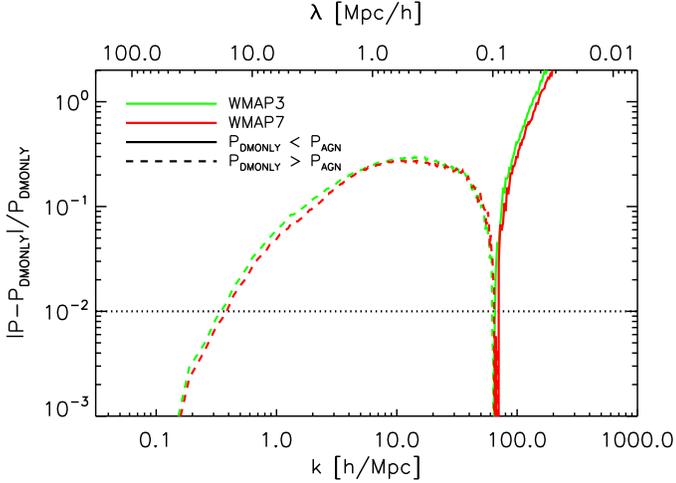}
\caption{The dependence of the effect of AGN on cosmology. The curves show the relative differences between the $z=0$ matter power spectra for models \textit{AGN} and \textit{DMONLY} for our fiducial WMAP3 cosmology (green) and for the WMAP7 cosmology (red). Changing the cosmology has little impact on the relative effect of the baryonic processes.}
\label{DMONLYAGN}
\end{center}
\end{figure}

The back-reaction is quite different when AGN feedback is included.\footnote{The small difference in power between the CDM component of \textit{AGN} and \textit{DMONLY} near the size of the box is most likely caused by errors in the power spectrum estimation.} The dark matter haloes still contract on small scales, albeit by a smaller amount, but the power in the dark matter component of the \textit{AGN} simulation is decreased for scales $> 200\lunitk$, corresponding to the sizes of haloes of $L_*$ galaxies. The reduction in the power of the CDM component in model \textit{AGN} relative to \textit{DMONLY} increases from roughly $1\%$ at $k=3\kunit$ to almost $10\%$ around $k=10\kunit$. AGN-driven outflows redistribute gas to larger scales, which reduces the baryon fractions in haloes and results in shallower potential wells. This is consistent with the results of \citet{Duffy2010}, who used the same simulation to show that AGN feedback decreases the concentrations of dark matter haloes of groups and clusters. Note, however, that because AGN can drive gas beyond the virial radii of their host haloes, their effect on the power spectrum cannot be fully captured by a simple rescaling of the halo concentrations. 

\subsection{A closer look at the effects of AGN feedback}
\label{sec:agn}
In this section we examine our most realistic model for the baryonic physics, \textit{AGN}, more closely. 
\subsubsection{Dependence on cosmology} 
Figure~\ref{DMONLYAGN} shows how the relative difference between the $z=0$ power spectra of models \textit{AGN\_WMAP7} and \textit{DMONLY\_WMAP7}, both of which use the WMAP7 cosmology, compares to that between the same physical models in the WMAP3 cosmology (the latter case was already shown in Figure~\ref{DMONLYREFAGN}). Even though the power spectra are themselves strongly influenced by, for example, the much higher value of $\sigma_8$ in the WMAP7 cosmology, the relative change in power due to baryons is nearly identical, at least so long as AGN feedback is included. This is good news for observational cosmology. It means that, once the large current scatter in implementations of subgrid physics has converged, it may be possible to separate the baryonic effects from the cosmological ones when modelling the matter power spectrum. It also means that we can assume that our results of the previous sections, which were based on the WMAP3 version of the \textit{AGN} simulation, apply also to model \textit{AGN\_WMAP7}.

\begin{figure}
\begin{center}
\includegraphics[width=0.95\columnwidth, trim=7mm 15mm 5mm -20mm]{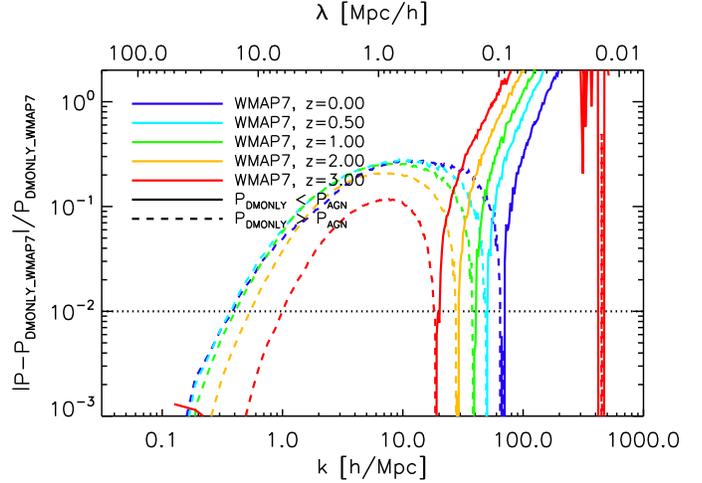}
\caption{Evolution of the relative difference between the matter power spectra of \textit{DMONLY\_WMAP7} and \textit{AGN\_WMAP7}. From red to blue, redshift decreases from $3$ to zero. The erratic behaviour of the $z=2$ and $z=3$ power spectra at the very smallest scales shown is due to a lack of resolution. For $\lambda \ga 1\lunit$ the reduction in power due to baryons evolves only weakly for $z\la 1$, but the transition from a decrease to an increase in power keeps moving to smaller scales.}
\label{DMONLYAGN_z}
\end{center}
\end{figure}

\subsubsection{Evolution}
Next, we use the \textit{AGN\_WMAP7} simulation, which we consider to be our most realistic model, to investigate the dependence of the effect of baryon physics on redshift. Figure~\ref{DMONLYAGN_z} shows the relative difference between the power spectra of \textit{DMONLY\_WMAP7} and \textit{AGN\_WMAP7} at redshifts $3$, $2$, $1$, $0.5$ and zero. We see from this plot that on large scales, $\lambda \ga 1\lunit$, the reduction in power due to the gas does not evolve much for $z\la 1$, although the differences between the different redshifts remain large compared with the precision of upcoming surveys. The weak evolution below $z=2$ is consistent with \citet{McCarthy2011}, who found that the expulsion of gas due to AGN feedback takes place primarily at $2 \la z \la 4$. On scales below $1\lunit$, on the other hand, the effects of baryonic processes on the power spectrum keep increasing with time, with the transition point between a decrease and an increase in power steadily moving towards smaller scales. This is probably because the ejection of low-entropy halo gas at high redshift ($z \ga 2$) results in an increase of the entropy, and thus a reduction of the cooling rates, of hot halo gas at low redshift \citep{McCarthy2011}.

\section{Comparison with previous work}
\label{sec:previouswork}
Our predictions for the effect of baryons on the matter power spectrum agree qualitatively with those of other authors, provided we restrict ourselves to including the same baryonic feedback processes as were considered in those studies. However, previous simulations did not include AGN feedback and hence suffered from overcooling.\footnote{The toy model of \citet{LevineGnedin2006}, which we briefly describe later in this section, did demonstrate, based purely on energetic grounds, that AGN feedback has the potential to have a large effect on the matter power spectrum.} As we have demonstrated, AGN feedback (or very efficient stellar feedback) has a dramatic effect on the matter power spectrum over a large range of scales. In this section we will consider both the qualitative and quantitative differences with respect to previous work, and examine how these may have come about.

\citet{Jing2006} used \textsc{gadget ii} \citep{Springel2005} to run a simulation with a $100\lunit$ box and $512^3$ gas and DM particles. Their simulation included radiative cooling and star formation, and used the \citet{SpringelHernquist2003} sub-grid model for the multiphase ISM and for galactic winds driven by star formation. Metal-line cooling and AGN feedback were not considered. They found that the power at $k=1\kunit$ is reduced by $\sim 1\%$ relative to a dark matter only simulation at $z=0$, which matches our results for the reference simulation very well. Furthermore, in agreement with our reference model, they find that the inclusion of baryons increases the power by $\sim 10\%$ at $k=10\kunit$. However, they find that the transition from a relative decrease to a relative increase in power occurs at $k \approx 2\kunit$, while we find that it lies at $k \approx 6\kunit$.

As the simulation of \citet{Jing2006} excludes metal-line cooling, we expect their results to be in better agreement with our own results for \textit{NOZCOOL}. The main difference with respect to the reference simulation turns out to be the position of the transition point from a relative decrease to a relative increase in power, which shifts to $k \approx 2-3\kunit$ when metal-line cooling is turned off. Hence, using the simulation \textit{NOZCOOL}, we reproduce both the qualitative and quantitative results of \citet{Jing2006}, even though baryonic processes such as SN feedback are not implemented in the same way.

\citet{Rudd2008} used the \textsc{art} code \citep{Kravtsov1999} expanded with the Eulerian hydrodynamics solver described in \citet{Kravtsov2002}. They used a $60\lunit$ box with $256^3$ particles, and included radiative cooling and heating, metal-line cooling, star formation, thermal SN feedback \citep[which is described in][]{Kravtsov2005} and chemical enrichment. AGN feedback was not considered. The effect of the baryons on the matter power spectrum they found is far more dramatic than that found by \citet{Jing2006} and ourselves: a decrease in power of up to $\sim 10\%$ relative to a dark matter only simulation for $k < 1\kunit$, and a relative increase in power at $k \ga 1\kunit$ which already reaches $\sim 50\%$ at $k \approx 5\kunit$. The reason for these large differences is unclear.

\citet{Guillet2009} used the MareNostrum simulation, which was run using the adaptive mesh refinement code \textsc{ramses} \citep{Teyssier2002}, to investigate the effects of baryons on both the variance and the skewness of the mass distribution. They used a $50\lunit$ box with $1024^3$ dark matter particles and included metal-dependent gas cooling, UV heating, star formation, SN feedback \citep[using the kinetic feedback prescription of][]{DuboisTeyssier2008} and metal enrichment. AGN feedback was not considered. Unfortunately, they were not able to run their simulations down to $z=0$, but quote results at redshift $2$ instead. In order to better compare to their results, we have examined the power spectra of \textit{REF\_L100N512} and \textit{DMONLY\_L100N512} at $z=2$. In our reference simulation the scale on which baryons significantly reduce the power increases with time (note that \textit{AGN} shows the opposite behaviour, see Fig.~\ref{DMONLYAGN_z}): in \textit{REF} the $1\%$ level is first reached at $k \approx 2\kunit$ for $z=2$ and at $k \approx 0.8\kunit$ for $z=0$. Meanwhile, the effect on the power on scales $k \ga 10\kunit$ hardly changes, and the transition scale from a decrease to an increase in power relative to \textit{DMONLY} remains fixed at $k \approx 7\kunit$. \citet{Guillet2009}, on the other hand, do not detect a systematic decrease in power due to baryons at any scale. They find that the power is increased by $1\%$ relative to a dark matter only simulation at $k \approx 3\kunit$, reaching $40\%$ at $k \approx 10\kunit$. For our reference model we instead find a $2\%$ decrease for $k \approx 3\kunit$ and only a $6\%$ increase at $k \approx 10\kunit$. It is hard to say why these results lie so far apart, and especially why the baryons in their simulation do not reduce the power on large scales due to pressure effects.

We also compare to the recent study by \citet{Casarini2010}, who use the SPH code \textsc{gasoline} \citep{Wadsley2004} to perform their simulations. They use two different volumes: a box of $64\lunit$ on a side, and a much larger $256\lunit$ box, both with only $256^3$ dark matter and an equal number of gas particles. Note that the mass resolution of their $L=64\lunit$ run is comparable to that of our fiducial run, while the resolution of their $L=256\lunit$ run is much poorer. They include radiative cooling, a UV background, star formation and SN feedback. For the latter they use the prescription of \citet{Stinson2006}, in which Type II SNe are modelled using an analytical treatment of blastwaves combined with manually turning off radiative cooling. Metal-line cooling and AGN feedback were not considered. Using their $64\lunit$ box, \citet{Casarini2010} find an $\sim 1\%$ decrease in power at $k \approx 1-2\kunit$ and an increase in power at smaller scales, which reaches $20\%$ at $k \approx 10\kunit$. These results are in reasonable agreement with both \citet{Jing2006} and our model \textit{NOZCOOL}. However, when using their $256\lunit$ box, they -- like \citet{Guillet2009} -- find no decrease in power due to baryons at any scale, but instead a steady increase in power that reaches $1\%$ at $k \approx 1-3\kunit$ and $40\%$ at $k \approx 10\kunit$.

Finally, we discuss the work by \citet{LevineGnedin2006}, who used a toy model, rather than a hydrodynamic simulation, to evaluate the potential effect of AGN feedback on the matter power spectrum. In their models only the evolution of dark matter was followed explicitly. The gas was assumed to trace the dark matter at all scales and galaxy formation and
the associated physical processes were not included. Their standard simulation volume is $64\lunit$ on a side, and the simulation was run with resolutions of $1$, $0.5$ and $0.25\lunit$. We note that even their highest resolution is more than two orders of magnitude below the spatial resolution in our standard simulations. The gas was assumed to have a constant temperature of $1.5 \times 10^4\,\mathrm{K}$ at all redshifts. A quasar luminosity function was used to determine the number of AGN at a given redshift and luminosity, which were then each placed at a random location, although biased towards high-density regions. Of the AGN's bolometric luminosity, a fraction $\epsilon_\mathrm{k}=1\%$ was used to drive spherically symmetric outflows. Within these outflow regions the baryon fraction was assumed to be zero. After computing the power spectrum, they found a large discrepancy between simulations with different resolutions: when using a resolution of $1\lunit$, they found a reduction of roughly $10\%$ in power for $0.3 \la k \la 3\kunit$ at $z=0$, relative to a simulation which did not include AGN, while their higher-resolution runs produced instead an increase in power at all scales, of up to $20\%$. We found a decrease in power of $1\%$ at $k \approx 0.3\kunit$, reaching $>10\%$ for $2 \la k \la 50\kunit$, which does not agree with their results, even in terms of the sign of the effect. Nevertheless, we do confirm the conclusion of \citet{LevineGnedin2006} that AGN feedback can greatly affect the matter power spectrum on a wide range of scales.

Even though our current understanding of galaxy formation still allows for significant deviations between studies, some qualitative results are the same: in the absence of AGN feedback, baryons will affect the matter power spectrum significantly on scales $k \sim 1-10\kunit$. Furthermore, all studies agree that the increase in power due to baryons is of the order of $10\%$ at $k=10\kunit$. \citet{Jing2006}, our reference and \textit{NOZCOOL} models, and \citet{Casarini2010} for their high-resolution simulation all predict a relative decrease in power of $\sim 1\%$ at $k \approx 1\kunit$. \citet{Rudd2008} also find a decrease in power due to baryons, but in their case the effect is far stronger than that of any other study, and is seen at much larger scales ($k \la 1\kunit$).

\emph{However, like our reference simulation, all these simulations suffer from the well-known overcooling problem.} As was demonstrated by \citet{McCarthy2010}, the \textit{AGN} simulation does not. We have shown that the inclusion of AGN has a tremendous effect on the matter power spectrum for $\lambda \la 10\lunit$, both when compared to a simulation that includes only dark matter and when compared to simulations that include baryons and galaxy formation but not AGN feedback. Therefore, contrary to what, for example, \citet{Guillet2009} claim, simulations that suffer from overcooling cannot be considered extreme models for which the effects of baryons on the total matter power spectrum are maximised. Instead, they are prone to \emph{under}estimate the effects on large scales. Indeed, model \textit{AGN} predicts a relative decrease in power of $\sim 1\%$ already at $k=0.4\kunit$. The decrease in power reaches several tens of percent on scales $k \sim 1-10\kunit$, while simulations that suffer from overcooling instead predict a strong increase in at least part of this range. Based on our results and on the comparison to other studies, we argue that the inclusion of AGN in cosmological simulations is at present even more important than the improvement or convergence of existing prescriptions for other baryonic effects.

Motivated by the results of \citet{Rudd2008}, \citet{Zentner2008} have proposed a method to account for the effects of galaxy formation on the matter power spectrum. This method assumes that the effects of baryons can be captured by a change in the halo concentration-mass relation. However, it is unlikely that such an approach can truly model the effects of baryons on the power spectrum. Since AGN-driven outflows significantly affect scales much larger than the sizes of individual haloes, the assumption made by \citet{Zentner2008} will certainly not be valid when AGN feedback is included.

\section{Conclusions}
\label{sec:conclusions}
Upcoming weak lensing surveys, such as LSST, EUCLID, and WFIRST aim to measure the matter power spectrum with unprecedented accuracy. In order to fully exploit these observations, theoretical models are needed that can predict the non-linear matter power spectrum at the level of $1\%$ or better on scales corresponding to $0.1\lunit \la k \la 10\kunit$. Here, we have employed a large suite of simulations from the OWLS project, as well as the highly accurate power spectrum estimator \textsc{powmes}, to investigate the effects of various baryonic processes on the matter power spectrum. These tools have also enabled us to examine the distribution of power over different mass components, the back-reaction of baryons on the CDM, and the evolution of the dominant effects on the matter power spectrum.

Our most important finding is that the feedback processes that are required to solve the overcooling problem (i.e.\ the overproduction of stars), have a dramatic effect on the matter power spectrum. Such efficient feedback, most likely in the form of outflows driven by AGN, were not present in the simulations used in previous studies of the effects of baryons on the matter power spectrum \citep{Jing2006, Rudd2008, Guillet2009, Casarini2010}. Although it was generally assumed that overcooling would make the simulations conservative, in the sense that they would overestimate the baryonic effects, we demonstrated that the opposite is true. The efficient outflows that are required to reproduce optical and X-ray observations of groups of galaxies, redistribute the gas on large scales, thereby reducing the total power by $\ga 10\%$ on scales $k\ga 1\kunit$. 

We emphasise that the model from which we draw this conclusion, the simulation that includes AGN feedback, is not extreme. On the contrary, we consider it our most realistic model. \citet{McCarthy2010,McCarthy2011} showed that it provides excellent agreement with both optical and X-ray observables of groups of galaxies at redshift zero. In particular, it reproduces the temperature, entropy, and metallicity profiles of the gas, as well as the stellar masses, star formation rates, and age distributions of the central galaxies, and the relations between X-ray luminosity and both temperature and mass. 

We showed that metal-line cooling, star formation, and feedback from SNe all modify the matter power spectrum by $>1\%$ on the scales relevant for upcoming surveys. In the absence of AGN feedback, the simulations with baryons have $\sim 1\%$ less power relative to a dark matter only simulation on scales $0.8 \la k \la 6\kunit$ (a consequence of gas pressure) and $>10\%$ more power for $k \ga 10\kunit$ (a consequence of gas cooling). However, as we noted above, AGN feedback can decrease the power for $1 \la k \la 10\kunit$ by up to several tens of percent. Furthermore, some implementations of stellar feedback, e.g.\ the strong SN feedback resulting from a top-heavy stellar initial mass function in starbursts, can create differences of the same scope and magnitude by redistributing gas out to very large scales. The effects from such baryonic processes on the matter power spectrum can even exceed those of a very large change in cosmology (e.g.\ WMAP3 to WMAP1). Indeed, differences $> 1\%$ persist even up to scales as large as those corresponding to $k \approx 0.3\kunit$. 

In the absence of AGN feedback, the back-reaction of baryons on the dark matter increases the power in the CDM component by $1\%$ at $k \approx 2\kunit$ and the effect becomes larger towards smaller scales. However, when AGN are included they redistribute sufficiently large quantities of gas out to large radii to lower the power in the dark matter component by $1-10\%$ for $3 \la k \la 30\kunit$. This  is consistent with \citet{Duffy2010}, who used the same simulation to show that AGN feedback decreases the concentrations of dark matter haloes of groups of galaxies. We stress, however, that the back-reaction of AGN feedback on the CDM will not be straightforward to implement in dark matter only models. While it may be possible to roughly model the effect of baryons in simulations without efficient feedback by raising the concentration parameters of the dark matter haloes \citep[e.g.][]{Zentner2008}, feedback from AGN redistributes the gas on scales that exceed those of their host haloes.

The difference between dark matter only simulations and simulations that do include baryons is nearly the same for the WMAP3 and WMAP7 cosmologies, at least when AGN are included. This suggests that the relative effect of the baryons is roughly independent of cosmology, which will simplify future studies aiming to disentangle the two. 

For our most realistic simulation, which assumes the WMAP7 cosmology and includes AGN feedback, the difference in power relative to the corresponding dark matter only simulation does not evolve much for $z\la 1$ on large scales ($k<10\kunit)$. This is consistent with \citet{McCarthy2011}, who showed that the expulsion of gas through AGN feedback occurs mostly at $z \sim 2- 4$, in the progenitors of today's groups and clusters of galaxies.

We demonstrated that our conclusions are robust with respect to changes in the size of the simulation box and changes in the resolution (see Appendix~A), with any additional modelling uncertainties only making it less likely that the matter power spectrum can be predicted with $1\%$ accuracy any time soon. Looking at the large differences that still exist between the results of different authors, it is clear that much work remains to be done in understanding processes such as gas cooling and outflows.

In a forthcoming paper (Semboloni et al., in preparation), we will study the implications of our findings for weak lensing surveys in more detail. In this work we will also demonstrate that the use of optical and X-ray observations of groups of galaxies can significantly reduce the uncertainties in the predictions of the matter power spectrum. While this provides a strong incentive for obtaining better and more observations of groups of galaxies, it is important to note that such auxiliary data will never completely remove the uncertainty inherent to cosmological probes of the matter distribution on scales that are potentially affected by baryonic physics. This is because one can never be sure that all the relevant effects are constrained by the secondary observations. For example, it may be that other models for the baryonic physics exist that also reproduce optical and X-ray observations of groups, but nevertheless predict different power spectra. It will therefore be crucial to consider a wide variety of observations, with optical and X-ray as well as Sunyaev-Zel'dovich observations holding particular promise, and a large range of models. 

While the strong baryonic effects that we find imply that the  cosmological constraints provided by upcoming weak lensing surveys will be model-dependent, it also means that such surveys will provide constraints on the physics of galaxy formation on scales that are difficult to obtain by other means.

Tabulated values of power spectra for redshifts $z=0-6$ are available for all the simulations shown in this paper as Supporting Information with the online version of this article, and at \texttt{http://www.strw.leidenuniv.nl/VD11/} (see Appendix~B).

\section*{Acknowledgements}
We thank all members of the OWLS team for their contributions to the project. We are also grateful to Henk Hoekstra, Elisabetta Semboloni and Simon White for discussions. We thank the Horizon Project for the use of their code \textsc{powmes}. Furthermore, in our comparison with analytical models we have utilized code from the Cosmology Initiative \textsc{iCosmo}. The simulations presented here were run on Stella, the LOFAR Blue Gene/L system in Groningen, on the Cosmology Machine at the Institute for Computational Cosmology in Durham as part of the Virgo Consortium research programme, and on Darwin in Cambridge. This work was sponsored by National Computing Facilities Foundation (NCF) for the use of supercomputer facilities, with financial support from the Netherlands Organization for Scientific Research (NWO). This work was supported by an NWO VIDI grant and by the Marie Curie Initial Training Network CosmoComp (PITN-GA-2009-238356).

\bibliographystyle{mn2e}
\bibliography{Marcelbib}

\appendix
\section{Convergence tests}
Here we investigate the effects of changing the box size or resolution of the reference simulation on its power spectrum.

\subsection{Box size}

\begin{figure}
\begin{center}
\includegraphics[width=0.95\columnwidth, trim=0mm -15mm 0mm -20mm]{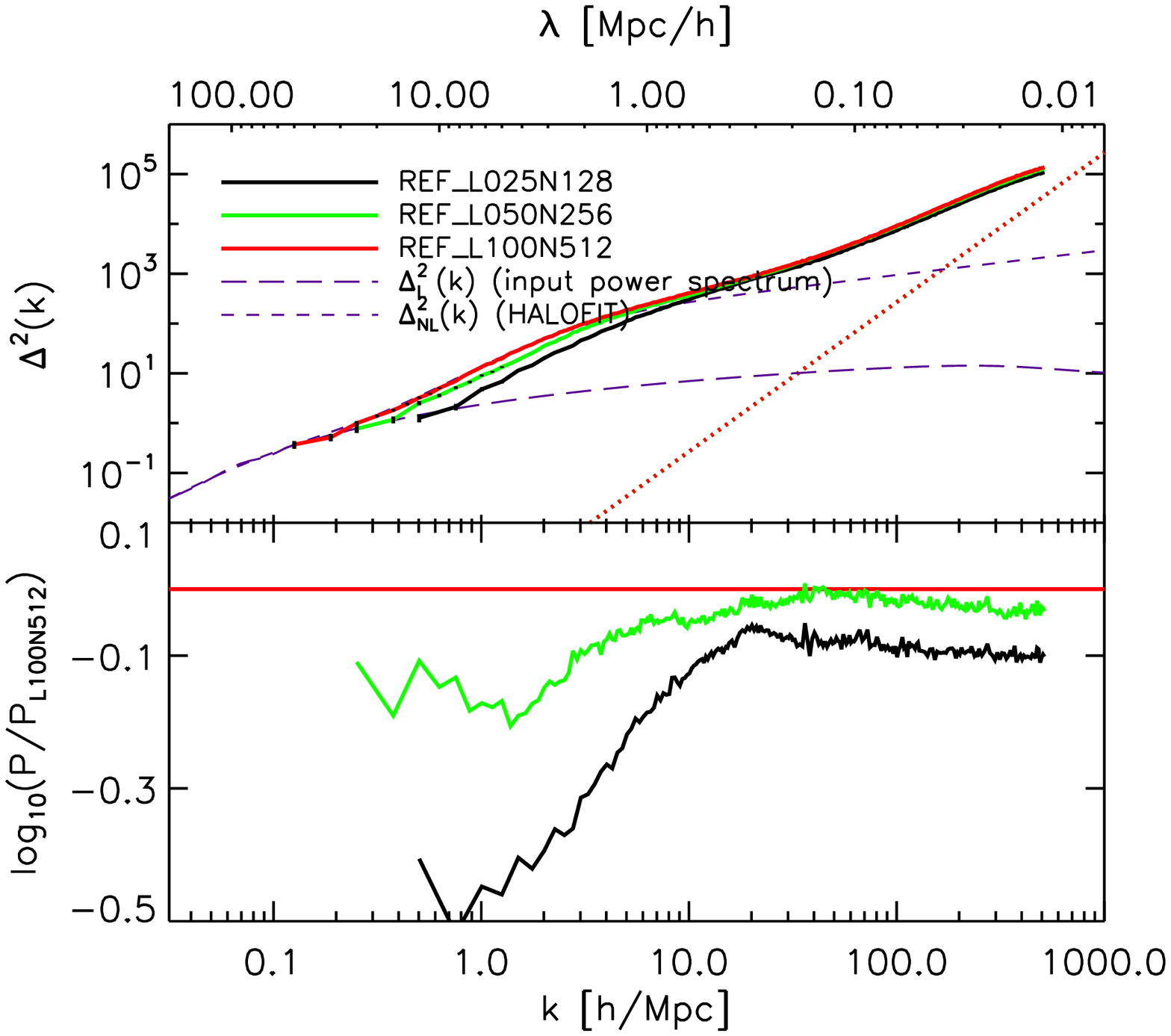}
\caption{Test of convergence of the $z=0$ matter power spectrum in the reference model with respect to the size of the simulated volume, where the box size and particle number are varied in such a way as to keep the resolution constant. Also shown are the linear input power spectrum and the analytical non-linear power spectrum by \citet{SmithPeacock2003}. The red, dotted line in the top panel shows the (subtracted) theoretical shot noise level. The bottom panel shows the ratio of \textit{REF\_L100N512} with respect to the other simulations.}
\label{restest_constres}
\end{center}
\end{figure}

In Figure~\ref{restest_constres} we vary the size of the box at constant resolution. The difference between the power spectrum of the 100 and the $50\lunit$ box is smaller than the difference between the latter and the $25\lunit$ box for all $k$, and the power spectrum of the largest box is nearly converged for $k \ga 20\kunit$.

However, there are differences of up to a factor of a few at larger scales. For reference, we also show the input power spectrum linearly evolved to $z=0$ and the HALOFIT model of the non-linear power spectrum by \citet{SmithPeacock2003} (see \S\ref{sec:dmonly}). The first wave mode corresponds to the size of the simulation box, which means that the power measured on this scale is meaningless; hence, we have omitted this point in all of our figures. The second and third wave modes closely follow the linear power spectrum. Note that the curves have very similar shapes on large scales, with the larger boxes shifted to larger scales. This is a consequence of employing the same seed for the random number generator used to create the initial conditions. Perturbations that should go non-linear ($\lambda \la 10\lunit$) are unable to collapse if their wavelength is close to the size of the box, which in turn suppresses the power on smaller scales. One might therefore worry that even the $100\lunit$ box is not large enough to obtain accurate power spectra for $k \ga 1\kunit$. Lacking larger simulations to check this, we compare to the HALOFIT model for the non-linear power spectrum, which shows where the transition from the linear power spectrum should take place at redshift zero. The power spectrum for the $100\lunit$ box follows this model very well on large scales, suggesting that a simulation of this size is very close to converged.

Note that finite volume effects only prevent us from obtaining highly accurate \textit{absolute} power spectra, and only for the largest scales, while our results are based on the \textit{relative} comparisons between models that used identical initial conditions. Since the $100\lunit$ box extends up to the largest non-linear scales, and since all simulations start from the \textit{exact same} realisation of the linear power spectrum at $z=127$, we do not expect our results to be affected by the finite volume of the simulations.

\begin{figure*}
\begin{center}
\begin{tabular}{ccc}
\includegraphics[width=0.95\columnwidth, trim=0mm -15mm 0mm -20mm]{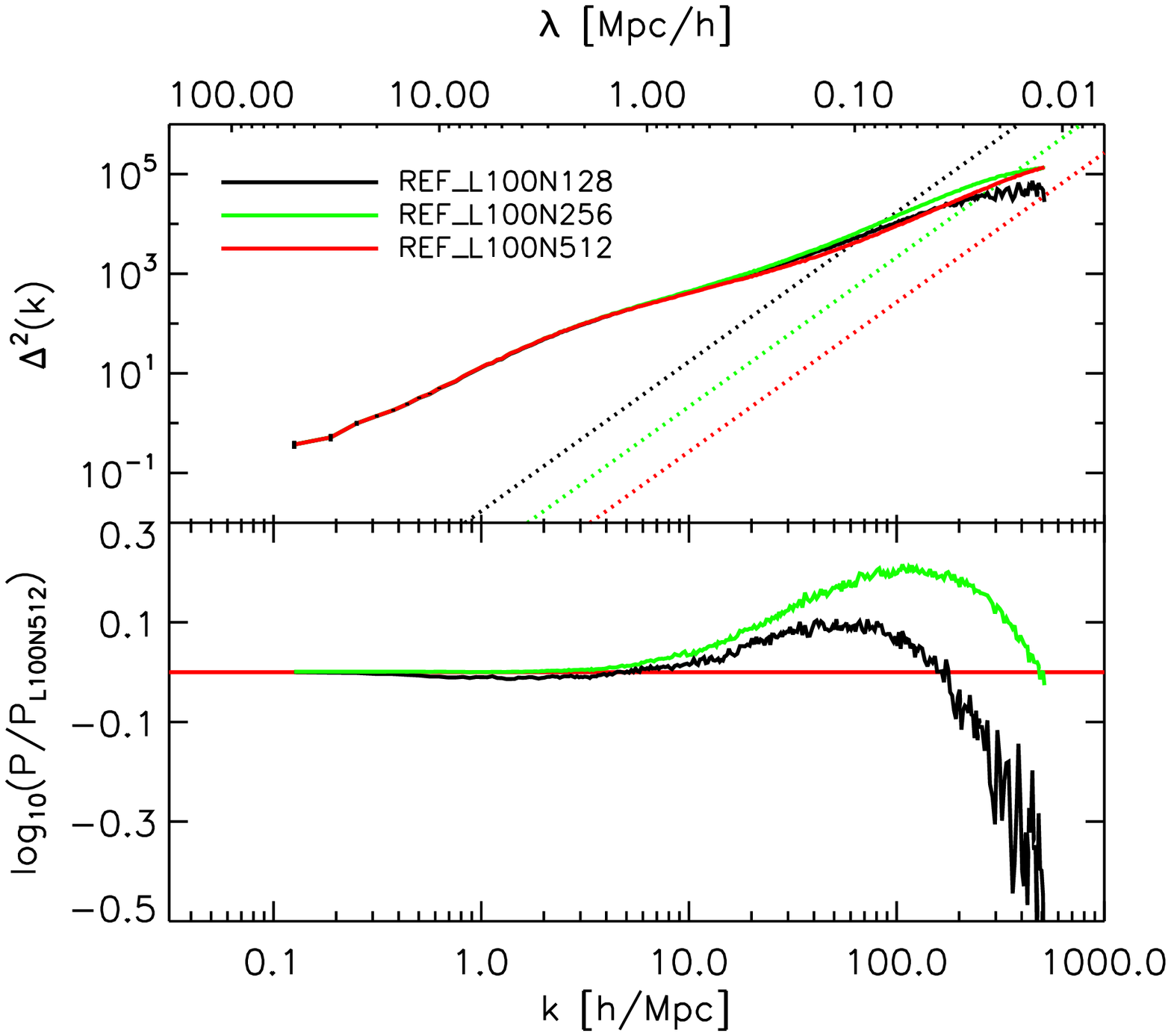} & &
\includegraphics[width=0.95\columnwidth, trim=0mm -15mm 0mm -20mm]{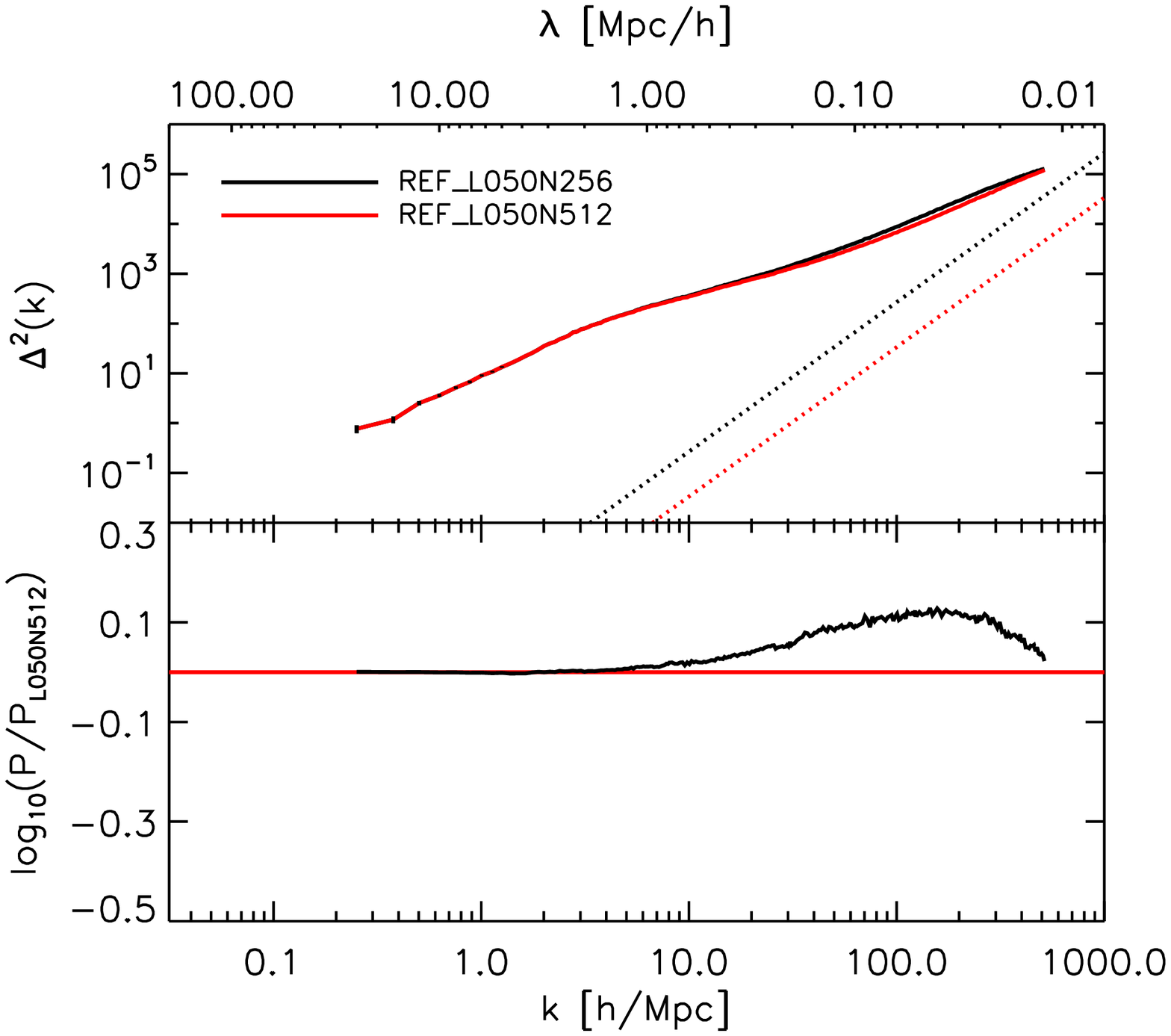}
\end{tabular}
\caption{As in Figure~\ref{restest_constres}, but now the numerical resolution is varied while keeping the box size constant. The left and right panels show power spectra for box sizes of 100 and $50\lunit$, respectively. On small scales, $k > 10\kunit$, convergence is slow.}
\label{restest_N}
\end{center}
\end{figure*}

\subsection{Numerical resolution}
In Figure~\ref{restest_N} we investigate the effects of changing the resolution for the reference simulation by varying the number of particles while keeping the box size fixed. The power spectrum of \textit{REF\_L100N128} is quite noisy for $k \ga 100\kunit$ because of its much higher Poisson noise level. Testing for convergence on these scales is only possible thanks to the accurate shot noise subtraction. Surprisingly, the relative difference in power between \textit{REF\_L100N128} and \textit{REF\_L100N512} is smaller than the difference between the latter and \textit{REF\_L100N256}. When increasing the resolution beyond that of \textit{REF\_L100N256}, the power begins to decrease. To examine if this trend continues, we compare the power spectrum of \textit{REF\_L050N256}, which has the same resolution as \textit{REF\_L100N512}, to that of \textit{REF\_L050N512} in the panel on the right. We see that the power on the smallest scales ($k\ga 10\kunit$) converges only slowly, but that the trend of decreasing power with increasing resolution continues. This may indicate that, as lower mass haloes become resolved, the overall effects of supernova feedback become stronger.

\begin{figure}
\begin{center}
\includegraphics[width=0.95\columnwidth, trim=0mm -15mm 0mm -20mm]{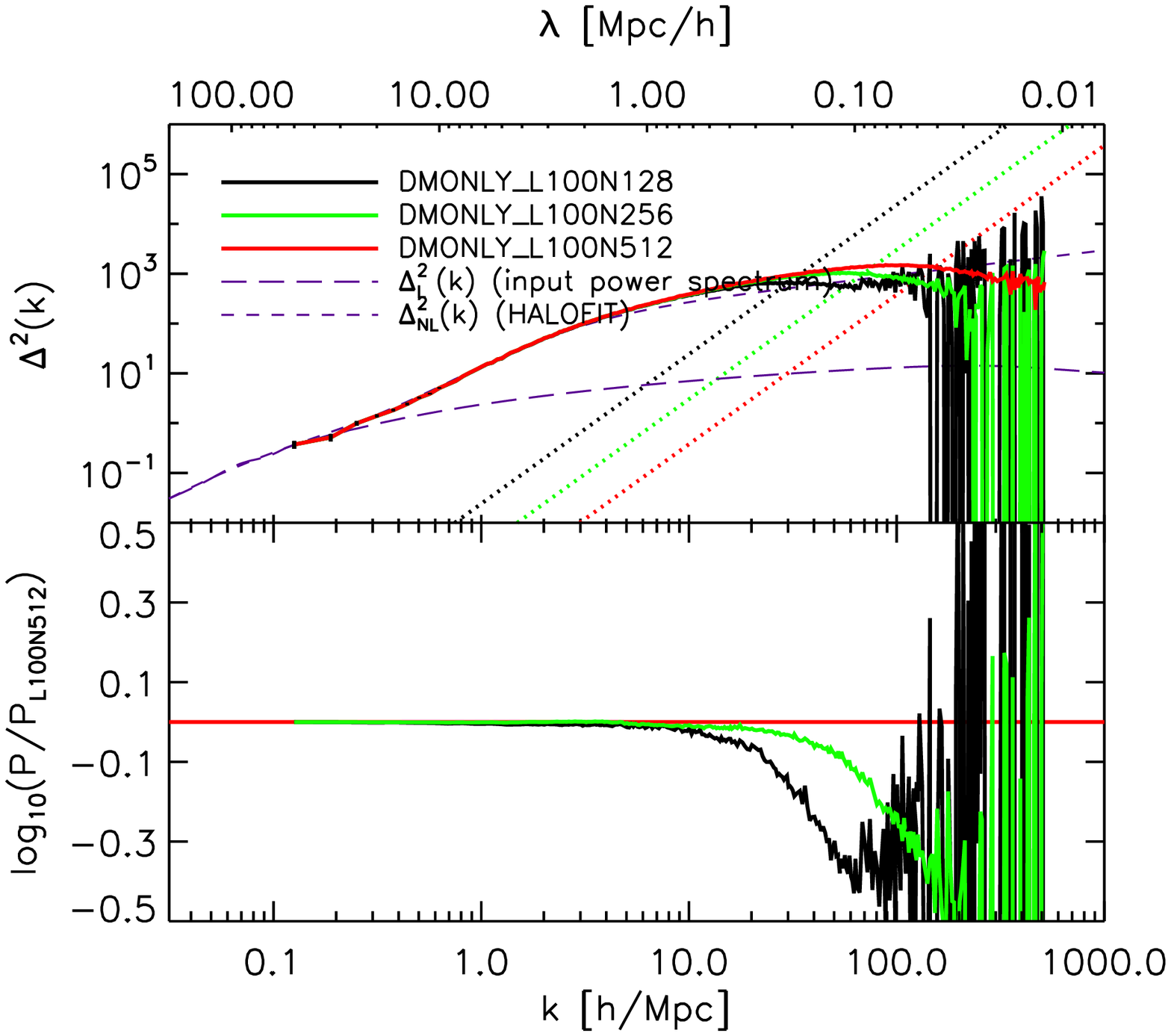}
\caption{Same as the left-hand panel of Figure~\ref{restest_N}, but now for \textit{DMONLY} instead of \textit{REF}. Here the behaviour is as expected: as the number of particles goes up, more low-mass haloes form and the power on small scales increases. A comparison with Figure~\ref{restest_N} shows that increasing the resolution leads to stronger baryonic effects which may reverse the sign of the trend with resolution.}
\label{restest_DMN}
\end{center}
\end{figure}

We can verify this by isolating the effects due to baryon physics from those due to a more straightforward dependence on resolution. To this end, we examine what the effect is of increasing the particle number of the \textit{DMONLY} simulation with a $100\lunit$ box in Figure~\ref{restest_DMN}. The behaviour here is quite different: as $N$ grows, more low-mass haloes are resolved and the power on small scales increases. As we observe a reversed trend in Figure~\ref{restest_N}, we conclude that the increased baryonic effects that accompany a higher particle number are more important for the power spectrum than the straightforward dependence on resolution.

The difference between \textit{REF\_L050N256} and \textit{REF\_L050N512} is $\sim 0.1\%$ at $k=1\kunit$ and $\sim 2\%$ at $k=10\kunit$. We conclude that simulation \textit{REF\_L100N512} is sufficiently converged for the scales of interest for this study, $k \la 10\kunit$. Note that, since we are only interested in the relative differences between simulations with equal resolution, the uncertainty will in practice be much smaller. With increased resolution we expect feedback processes to become more effective, meaning that we may have underpredicted the differences between models with different feedback processes in low-mass haloes on small scales.

Similar tests were performed by \citet{Colombi2009} for the convergence of \textsc{powmes}, which keeps the statistical error bounded through its use of foldings. Its value depends on the quantity $C(k)$, which is defined as the number of independent wave modes at a given wave number $k$; to be more precise, we approximately have $\Delta P/P \propto C(k)^{-1/2}$ \citep{Colombi2009}. For our fiducial grid with $256^3$ grid cells, one can expect the statistical error to remain below $|\Delta P|/P \approx 1.2\%$ as long as errors due to shot noise do not dominate. We have checked that this is indeed the case. Note that this means that we can confidently measure $1\%$ differences between simulations using our fiducial values, as we are interested in systematic offsets covering at least a small range of scales in $k$-space, rather than random deviations.

\section{Tabulated power spectra}
Table~\ref{powtable} shows the power spectrum values for our most current and realistic simulation to date, \textit{AGN\_WMAP7\_L100N512}, for a subset of scales at $z=0$. Our fiducial \textsc{powmes} values of $256^3$ grid points and $7$ foldings were used, and shot noise has been subtracted. The full table, with power spectrum values at all scales shown in this paper and redshifts up to $z=6$, as well as tabulated data for all other simulations presented in this paper, are available as Supporting Information with the online version of this article and at
\texttt{http://www.strw.leidenuniv.nl/VD11/}.

\begin{table}
\caption{Power spectrum values for \textit{AGN\_WMAP7\_L100N512} for a subset of scales at $z=0$ (full table available online).}
\centering
\begin{tabular}{| c c | c c | c c | c c |}
\hline\hline
 & $z$ & $k\,[h/\mathrm{Mpc}]$ & $P(k)\,[h^{-3}\mathrm{Mpc}^3]$ & $\Delta^2(k)$ & \\ [0.5ex]
\hline
 &      0.000 &      0.12566371 &       4364.4776 &      0.43876514 & \\
 &      0.000 &      0.18849556 &       1853.4484 &      0.62886024 & \\
 &      0.000 &      0.25132741 &       1524.3814 &       1.2259802 & \\
 &      0.000 &      0.31415927 &       1112.5603 &       1.7476056 & \\
 &      0.000 &      0.37699112 &       847.62970 &       2.3007519 & \\
\hline
\end{tabular}
\label{powtable}
\end{table}

\bsp
\label{lastpage}
\end{document}